\documentclass[apj]{emulateapj}
\pdfoutput=1
\usepackage{booktabs}
\usepackage{graphicx}
\usepackage{amssymb}
\usepackage{url}
\usepackage{hyperref}
\hypersetup{pdfdisplaydoctitle=true,hyperfootnotes=false}
\usepackage{breakurl}
\usepackage{color}
\usepackage[section]{placeins} 
\def\sun{\hbox{$_\odot$}}
\def\G48{G48}
\def\13CO{\ce{^{13}CO}}
\def\C18O{\ce{C^{18}O}}
\def\H2{\ce{H2}}
\def\H2O{\ce{H2O}}
\usepackage{natbib}
\bibpunct{(}{)}{;}{a}{}{,} % to follow the A&A style
\usepackage[version=3]{mhchem}

\begin{document}

\title{G048.66$-$0.29: Physical State of an Isolated Site of Massive Star Formation}
\shorttitle{IRDC G048.66$-$0.29}
\author{J.Pitann\altaffilmark{1}, H.Linz\altaffilmark{1}, S.Ragan\altaffilmark{1}, A.M.Stutz\altaffilmark{1}, H.Beuther\altaffilmark{1}, Th.Henning\altaffilmark{1}, O.Krause\altaffilmark{1}, R.Launhardt\altaffilmark{1},  A.Schmiedeke\altaffilmark{3}, F.Schuller\altaffilmark{4}, J.Tackenberg\altaffilmark{1}, T.Vasyunina\altaffilmark{2}}
\affil{$^1$Max-Planck-Institut f\"ur Astronomie (MPIA), K\"onigstuhl 17, D-69117 Heidelberg, Germany,\\ 
$^2$Department of Chemistry, University of Virginia, Charlottesville, VA 22904, USA\\
$^3$Physikalisches Institut, Universit\"at zu K\"oln, Z\"ulpicher Str. 77, D-50937 K\"oln, Germany\\
$^4$Max-Planck-Institut f\"ur Radioastronomie (MPIfR), Auf dem H\"ugel 69, 53121, Bonn, Germany
}
\email{pitann@mpia.de}
\shortauthors{Pitann et al. (2012)}
\submitted{}
\journalinfo{Submitted to \apj:  September 19, 2012, Accepted for publication}

\begin{abstract}
We present continuum observations of the infrared dark cloud (IRDC)
G48.66–0.22 (\G48) obtained with {\it Herschel}, {\it Spitzer}, and
APEX, in addition to several molecular line observations.  The {\it
Herschel} maps are used to derive temperature and column density maps
of \G48 using a model based on a modified blackbody. We find that \G48 has a relatively simple
structure and is relatively isolated; thus this IRDC provides an excellent target to study the
collapse and fragmentation of a filamentary structure in the absence of complicating factors such as strong external feedback.  The derived temperature structure of \G48 is clearly non-isothermal from cloud to
core scale. The column density peaks are spatially coincident with the lowest
temperatures ($\sim17.5\,$K) in \G48. A total cloud mass of $\sim 390\mathrm{M}_\odot$ is derived from the column density maps.  By comparing the luminosity-to-mass ratio of 13 point sources detected in the \textit{Herschel}/PACS bands to evolutionary models, we find that two cores are
likely to evolve into high-mass stars ($M_\star \geq 8\,\mathrm{M}_\odot$). The derived mean projected  separation of point sources is smaller than in  other IRDCs but in good agreement with theoretical predications for
cylindrical collapse.  We detect several molecular species such as \ce{CO},
\ce{HCO^+}, \ce{HCN}, \ce{HNC} and \ce{N2H^+}. CO is depleted by a factor of $\sim 3.5$ compared
to the expected interstellar abundance, from which we conclude that \ce{CO}
freezes out in the central region. Furthermore, the molecular clumps,
associated with the sub-millimeter peaks in \G48, appear to be
gravitationally unbound or just pressure confined. The analysis of critical line masses in \G48
show that the entire filament is collapsing, overcoming any internal
support. 

\end{abstract}

\maketitle

\section{Introduction}
\label{Intro}

\addtocounter{footnote}{2}
\footnotetext[1]{\textnormal{{\it Herschel}  is an ESA space observatory with science instruments provided by European-led Principal Investigator consortia and with important participation from NASA.}} 
\footnotetext[2]{Based on observations carried out with the IRAM 30m Telescope. IRAM is supported by INSU/CNRS (France), MPG (Germany) and IGN (Spain).}

%\clearpage
%\begin{landscape}
\begin{figure*}
	\centering
	\includegraphics[scale=0.6,angle=270]{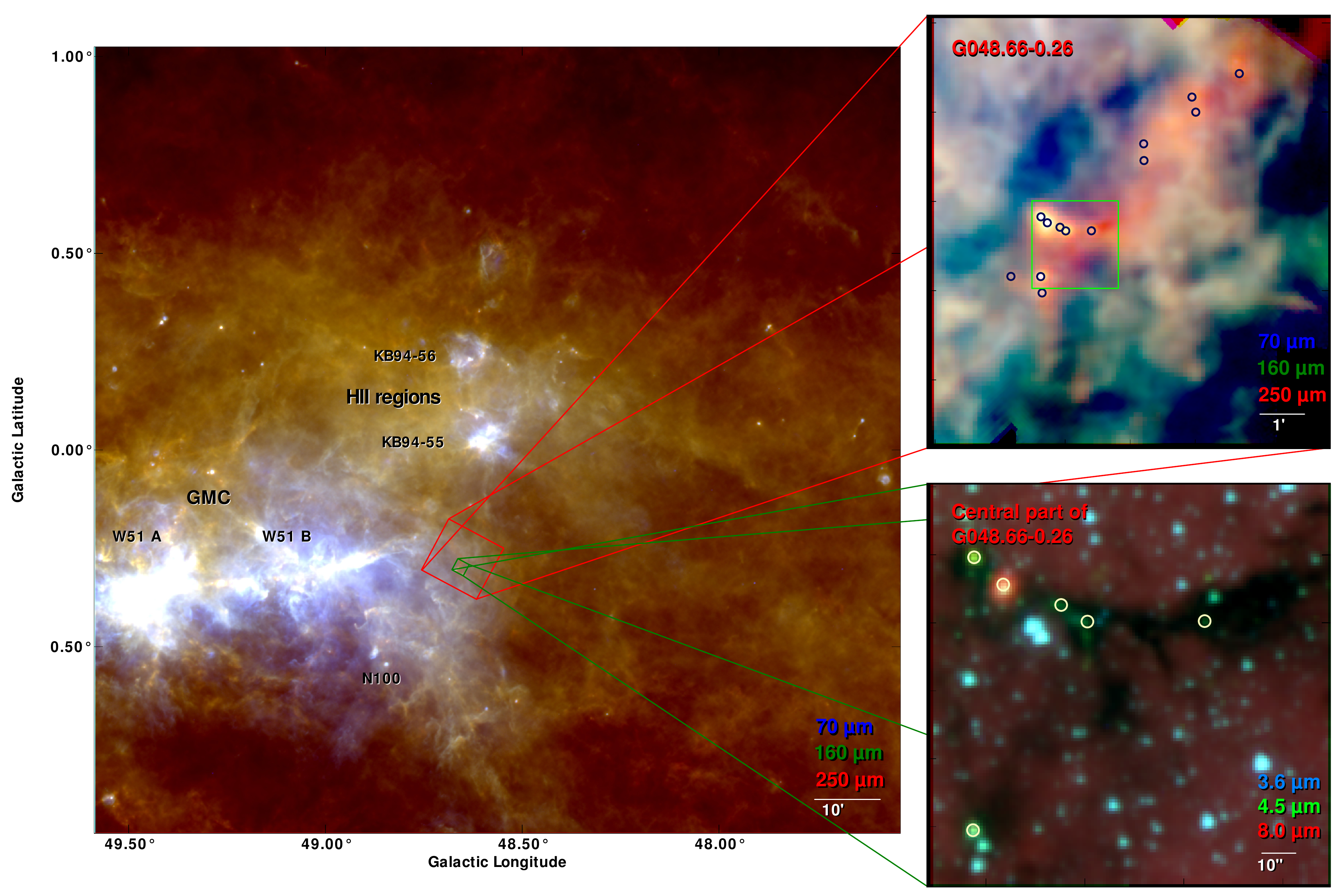}
	\caption{The \textit{upper panel} presents a \textit{Herschel} map of the galactic neighborhood of \G48. The dominant feature to the left of \G48 is the Giant Molecular Cloud complex of W51. Two HII regions described by \citet{Kuchar94} are located above \G48. Below \G48 the infrared bubble N100 appears.
	The map is composed of the PACS band at 70\,$\mu$m (blue), 160\,$\mu$m (green) and SPIRE at 250\,$\mu$m (red) from the Hi-GAL field 48\_0. The \textit{bottom right panel} is a zoom into the whole IRDC. It uses the same \textit{Herschel} bands, utilizing the EPOS observations. The \textit{bottom left panel} shows the central part of \G48 as an IRAC 3-color image (3.5, 4.8, 8.0\,$\mu$m). The green band shows elongated lobes attributed to shock-excited H$_2$, sometimes refereed to as ``green and fuzzy'' features ``or extended green objects'' (EGOs). The detected \textit{Herschel} point-sources from Section \ref{sec:pointsrc} are indicated by circles in the right panels. All intensities are plotted with logarithmic scaling in all maps.}
	\label{fig:RGB_overview}
\end{figure*}
%\end{landscape}
%\clearpage

%%%%%%%%%%%%%%%%%%
%                %
%   Begin Text   %
%                %
%%%%%%%%%%%%%%%%%%

Our knowledge of the physical conditions during the very early phases of high-mass star formation ($\gtrsim 8\mathrm{M}_\odot$) has been relatively limited in the past \citep[e.g.,][]{Beuther07}, due to significant observational challenges.
Consequently, the qualitative and quantitative differences in star-formation processes over different mass regimes are still strongly debated \citep[cf.][]{Zinnecker07}.
In the endeavor to enhance our understanding of star formation beyond the well-studied low mass stellar nurseries, infrared dark clouds (IRDCs) have become more and more important targets. Initially IRDCs were discovered by the Infrared Space Observatory \citep[ISO, ][]{Perault96} and the Midcourse Space Experiment \citep[MSX, ][]{Egan98}. 
They appear as filamentary opaque structures against the diffuse galactic mid-infrared background. As IRDCs are known to be cold ($\lesssim 20$\,K), dense, and exhibiting high column densities $(\gtrsim 10^{23}$cm$^{-2})$, they are suitable environments to form young stellar objects that can evolve into stars of intermediate to high mass \citep[e.g., ][]{Carey00,Rathborne06,Ragan06,Ragan09,Vasyunina09}. IRDCs were widely discussed in the context of massive star formation, but a certain fraction of IRDCs only host sub-solar to intermediate-mass star formation \citep{Beuther07,Kauffmann10,Peretto10a}. Still, \citet{Rathborne06} argue that most stars in our galaxy could form in IRDC-like molecular structures. 

Most of the current theories about the formation of IRDCs treat the collapsing gaseous medium as isothermal. 
With the \textit{Herschel} observatory, we have the spatial resolving power, spectral coverage, sensitivity and dynamic range to study the temperature and column density structure of IRDCs in more detail. 
Already one of the early \textit{Herschel} studies using spatially resolved SED fitting toward an IRDC indicated a non-isothermal temperature structure \citep{Peretto10}.

\citet{Wilcock12} compared \textit{Herschel} observations of IRDCs without embedded IR sources to radiative transfer models of strictly externally heated structures. They find resulting temperature profiles that go down to 11.5\,K (sample average) in the center. 
Two further studies analyzed the temperature structure using SED-fitting and a modified blackbody model.
\citet{Battersby11} present large scale temperature maps of IRDCs. \citet{Beuther11} studied the IRDC18454 near the W43 mini-starburst and present dust temperature maps. Both studies found temperatures higher by more than 5\,K compared to the IRDCs from \citet{Wilcock12}.
Further investigations have to clarify if the majority of IRDCs show similar minimum temperatures or if clumps with higher central temperatures exist. This classification is essential for the understanding of the formation process of massive stars. 
The comparison between dust and gas temperatures can shed light on the efficiency of the gas-dust coupling at densities typical for IRDCs. Furthermore, a well constrained dust temperature structure can provide critical input for 
chemical models of IRDCs, and in particular for the dust grain surface reactions. 

The goal of this paper is to accurately characterize the physical conditions of (high-mass) star formation in a relatively isolated IRDC, G048.66$-$0.29 (henceforth called ``\G48''). This case study is based on observations taken as part of the Earliest Phases of Star Formation (EPoS) Herschel guaranteed time key program \citep{Launhardt12,Ragan12}. We present sub-parsec scale dust temperature and column density maps and place these properties in context with the kinematic environment, probed by various molecular gas tracers (\ce{CO}, \ce{HCO^+}, \ce{N2H^+}, \ce{HCN}, \ce{HNC}). Herschel also enables us to characterize the embedded protostellar core population and, through the use of evolutionary models, try to understand the likely outcome of star formation in this cloud. 

The kinematic distance to \G48 was determined to be $\sim 2.7$\,kpc \citep{Ormel05} 2.5\,kpc \citep{Simon06a} or 2.6\,kpc (this work).
\G48 appears to be projected near the GMC W51 (Figure \ref{fig:RGB_overview}, left), which is in fact twice as distant as \G48 \citep[][see also Section \ref{sec:dist}]{Sato10}.
\G48 appears isolated; and therefore we can assume it is not strongly influenced by local heating sources.
Furthermore \G48 has a quite simple monofilamentary structure. By comparing the observational results from this isolated IRDC to more complex IRDCs \citep[c.f.][]{Ragan12}, we can constrain the dominant effects at work for star formation in IRDCs.
\citet{Simon06a} estimated from \13CO observations an average density for H$_2$ associated with \G48 of ${\sim 3 \cdot 10^{3}\,\textrm{cm}^{-3}}$, a peak column density of ${8.8\cdot10^{21}\,\textrm{cm}^{-2}}$ and a LTE mass of ${\sim580\,\textrm{M}_\odot}$.
\citet{Ormel05} identified three different clumps with masses from 91 to ${130\,\textrm{M}_\odot}$ using  SCUBA and MSX observations.  
By modeling the internal velocity structure of these clumps using \ce{HCO^+} $(3-2)$ and $(4-3)$ single-dish data  they found that the line-widths decrease outside the cloud, indicating a decrease in turbulent motions outwards of the clumps. 
This suggests that the star formation activity inside these clumps down to the core scale is driven by turbulence.

\citet{Wiel08} identified 13 YSOs in the vicinity of \G48, based on \textit{Spitzer} mid-infrared
maps. The majority of these objects belong to ``Stage I'', referring to the classification scheme of \citet{Robitaille06}, and have low to intermediate masses. 
From a comparison of [CI] and \ce{^{13}CO} data, \cite{Ossenkopf11} inferred the presence of a young photodissociation region (PDR) ionized by the YSOs inside of \G48.

Throughout the literature the limits of sizes and masses for clumps and cores vary.
In this work the spatial definition from \citet{Bergin07} is applied: cores range from $0.03-0.2$\,pc, while clumps extend over $0.3-3$\,pc.

Given the isolated nature and simple structure of \G48, we use this source to study the collapse and fragmentation of a filamentary IRDC. 
Through the characterization of the point sources the evolutionary stage can be assessed a final stellar mass can be estimated from different models. 
Furthermore, the \textit{Herschel} maps are utilized to study the temperature and column density structure of \G48. Molecular line observations provide information about the morphology of the clumps and clues to the pressure support therein. 
We assess the level of freeze-out by studying the CO-depletion in the central part of the IRDC.

The next section describes the observations, the data reduction and some basic data processing steps for \textit{Herschel}, \textit{Spitzer},  sub-millimeter continuum and molecular line observations. 
Section \ref{sec:Results} details the scientific analysis and the direct results. 
In Section \ref{sec:morpho} the structure and environment of \G48 is described.
The measured distances for \G48 and other sources in the near projected distance are discussed in Section \ref{sec:morpho}.
The temperature mapping, utilizing \textit{Herschel}, SCUBA and ATLASGAL data, is delineated in Section \ref{sec:Tmap}. 
The subsequent Section \ref{sec:Colmap} compares the column densities with other observations and estimates mass for the whole IRDC. 
Section \ref{sec:pointsrc} addresses the fitting and classification of point sources from the MIPS 24\,$\mu$m and PACS bands. 
Section \ref{sec:molecline} contains the analysis of the molecular line data. 
The concluding Section \ref{sec:conclusion} discusses the results in respect to previous studies of IRDC and theoretical considerations. 
The last Section \ref{sec:summary} summarizes the main results.

\section{Observations, Data reduction and processing}
\begin{figure}
	\centering
	\plotone{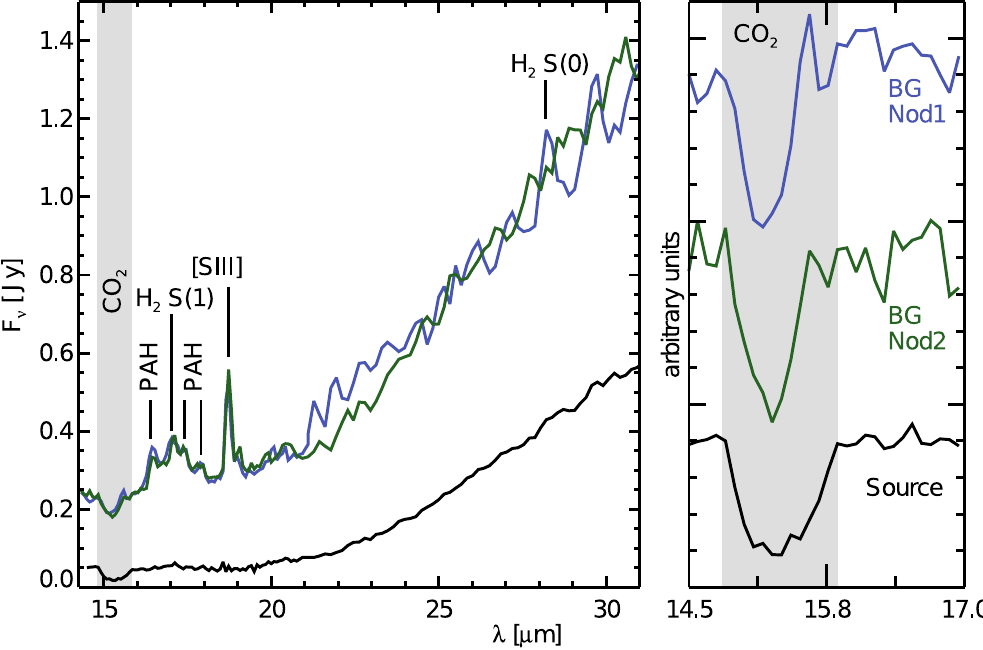}
	\caption{The LL-spectrum from Source 2 from IRS/\textit{Spitzer} is shown. In the left panel the source spectrum is plotted in black and the extracted background spectra are shown in blue and green. The background spectrum extracted in the first nod position (BG Nod 1) pointed towards the GMC, while the second background spectrum is taken at the western edge of the IRDC (BG Nod 1).  The atomic [SIII] line is present as a bright extended emission over the whole slit. The CO$_2$ ice absorption band is underlined in light grey. The right plot shows the baseline subtracted CO$_2$ profiles. }
	\label{fig:spec}
\end{figure}

\begin{figure*}
	\centering
	\includegraphics[width=\textwidth]{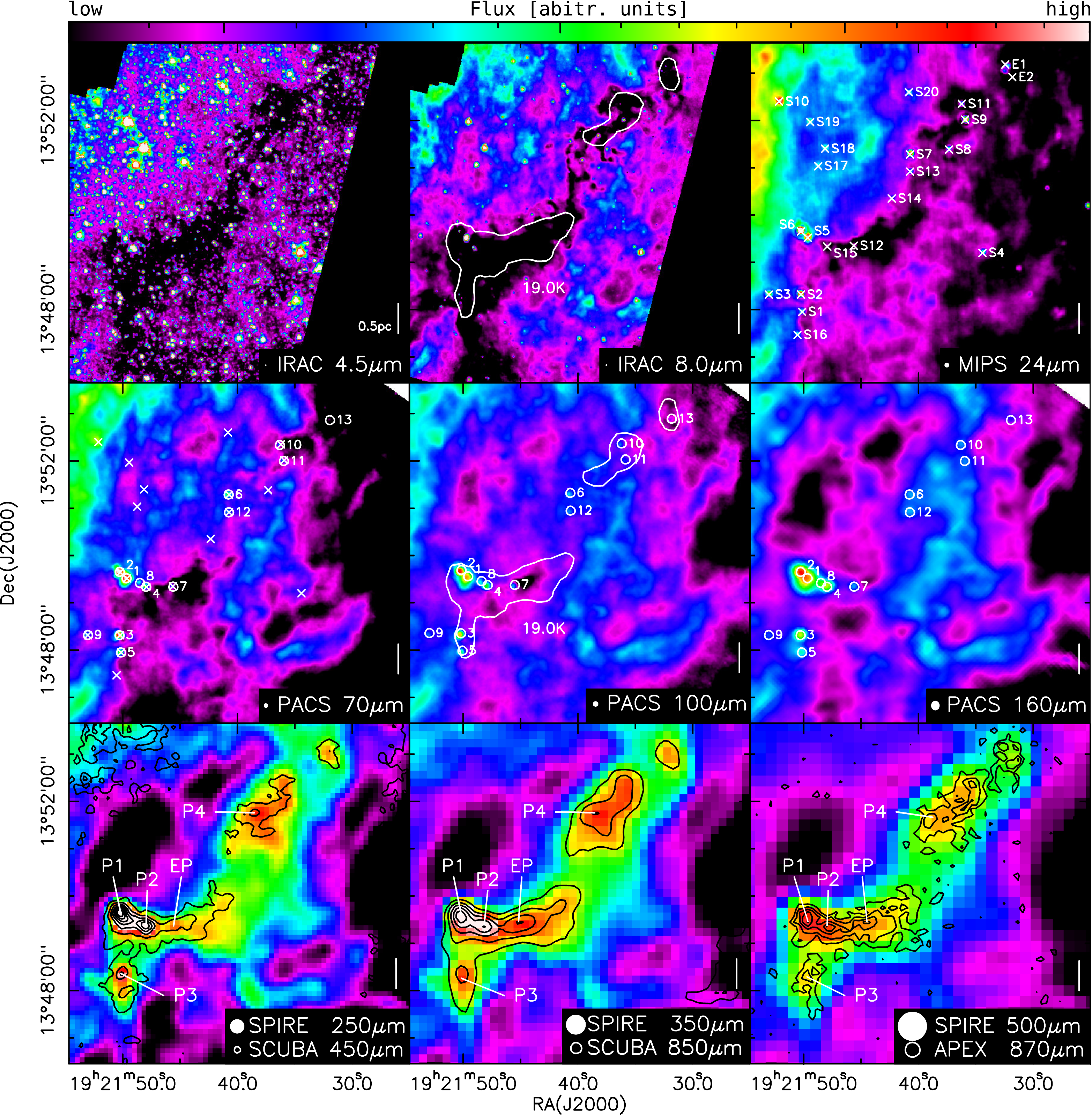}
	\caption{The intensity plots for \G48 are shown in the different \textit{Spitzer}, PACS and SPIRE channels. Each plot is shown on a linear scale to highlight absorption or emission at each wavelength band; the top color bar indicates the relative flux scale.
	The positions of the 24\,$\mu$m point sources as characterized by \citet{Wiel08} are indicated by red crosses in the MIPS 24\,$\mu$m and PACS 70\,$\mu$m maps. The MIPS~24\,$\mu$m sources from \citet{Wiel08} are labeled ``Sx'' while two additional 24\,$\mu$m sources associated to \G48 are labeled ``E1'' and ``E2''.	The detected sources in the PACS bands are marked by the white circles. The radius of the circles imply the fitting radius between the different bands. 
	The SPIRE maps are overlayed by SCUBA 450\,$\mu$m, 850\,$\mu$m and ATLASGAL 850\,$\mu$m contours, respectively. The corresponding beam sizes for \textit{Spitzer}, PACS and SPIRE are indicated with circles in each panel.
	An isothermal contour of 19\,K from the temperature fitting (Section \ref{sec:Tmap}) is overplotted for IRAC 8\,$\mu$m and PACS 100\,$\mu$m.
	The beam size of the observations plotted in contours are plotted as open ellipses. The white bar (right side) gives a projected scale of 0.5\,pc on the map. }
	\label{fig:stamps}
\end{figure*}
\begin{figure*}
	\centering
	\includegraphics[height=0.95\textheight]{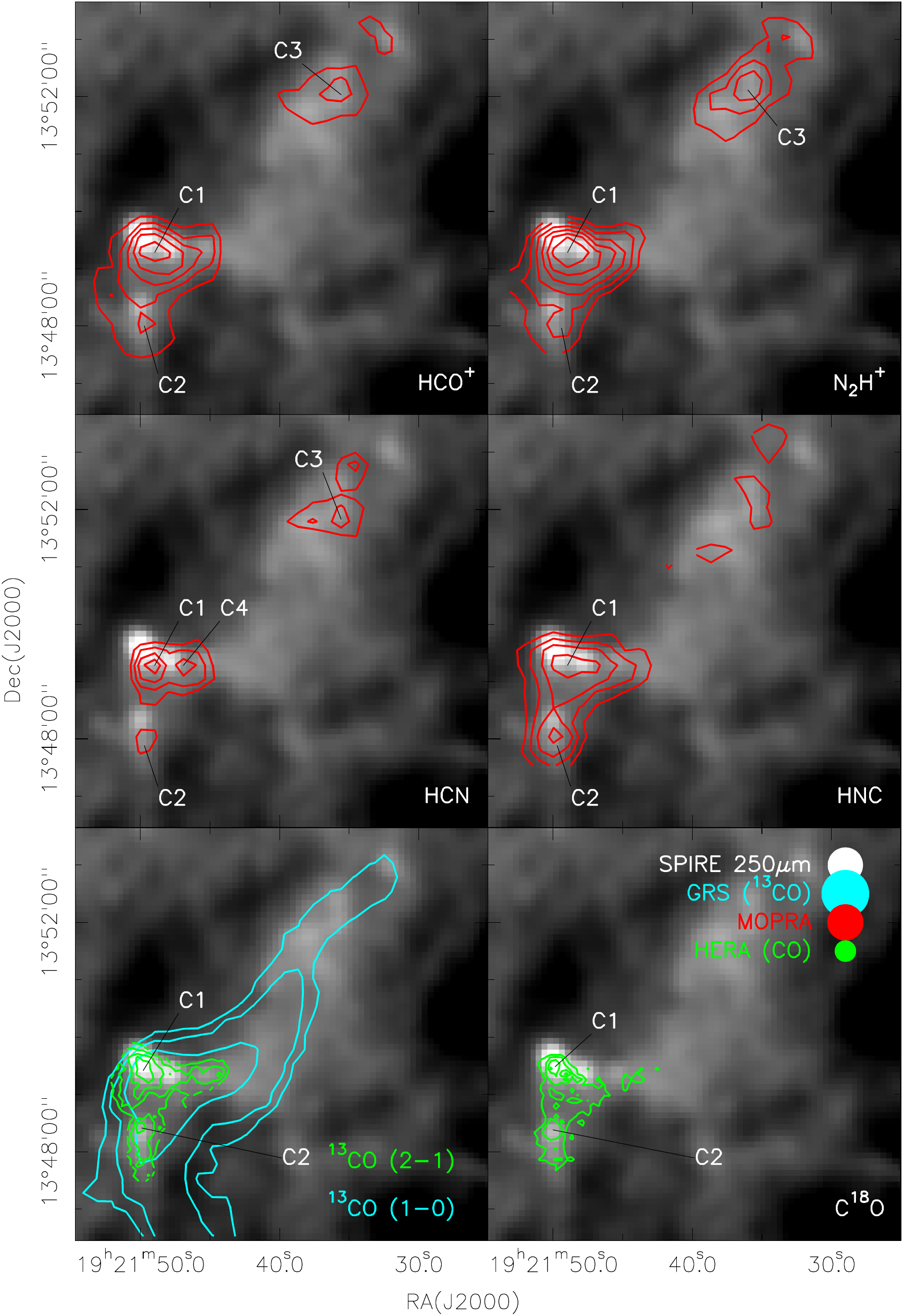}
	\caption{The integrated line intensities of the indicated species are shown as contours overlaying the 250\,$\mu$m SPIRE map. The contours in the top and middle panels refer to the MOPRA observations, while the bottom panels present the CO data. The beam sizes are indicated in the last panel.
The maps are integrated between 32 and 36\,km/s. The contours for the different line transitions are:\newline
		\ce{HCO^+}:	from 0.3 to 1.5\,K\,km\,s$^{-1}$ in steps of 0.3\,K\,km\,s$^{-1}$; \ce{HCN}: from 0.2 to 0.8\,K\,km\,s$^{-1}$ in steps of 0.1\,K\,km\,s$^{-1}$; \newline
		\ce{HNC}: from 0.2 to 0.8\,K\,km\,s$^{-1}$ in steps of 0.1\,K\,km\,s$^{-1}$; \ce{N2H^+}: from 0.5 to 2.0\,K\,km\,s$^{-1}$ in steps of 0.25\,K\,km\,s$^{-1}$; \newline
		\ce{^{13}CO}\,(1-0): from 0.8 to 2.9\,K\,km\,s$^{-1}$ in steps of 0.7\,K\,km\,s$^{-1}$; \ce{^{13}CO}\,(2-1): from 2.0 to 9.5\,K\,km\,s$^{-1}$ in steps of 1.5\,K\,km\,s$^{-1}$;\newline
		\ce{C^{18}O}\,(2-1): from 0.7 to 2.7\,K\,km\,s$^{-1}$ in steps of 0.5\,K\,km\,s$^{-1}$. } 
	\label{fig:mopra}
\end{figure*}

\subsection{Herschel}

In this paper we present results from the \textit{Herschel} GTO key program - ``Early Phases of Star Formation'' (EPoS). The high-mass portion of EPoS is compiled from different surveys. The sample includes a sub-sample of
``classical'' IRDCs, a selection of IRDCs in the near vicinity of high-mass proto-stellar objects (HMPOs) and cold, massive, compact sources detected by the ISO Serendipity Survey (ISOSS). For a detailed discussion of this sample  and the underlining selection strategy we refer the reader to the overview paper by \citet{Ragan12}. 

The EPoS observation were supplement with data from the the "Herschel infrared Galactic Plane Survey" (Hi-GAL) project, done with the Photometer Array Camera and Spectrometer  \citep[PACS, ][]{Poglitsch10} and the Spectral and Photometric Imaging Receiver \citep[SPIRE, ][]{Griffin10}  on board of the \textit{Herschel Space Observatory} \citep{Pilbratt10}. 

\subsubsection{PACS}
The EPoS PACS scan maps at 70\,$\mu$m, 100\,$\mu$m and 160\,$\mu$m were obtained with the medium scan speed of $20''$/s and the nominal scan leg length of 7$'$. The data were processed to level-1, i.e. calibrated and converted to physical units, using
HIPE \citep{Ott10}. We applied a second level deglitching to flag bad data values by $\sigma$-clipping flux values mapped on to each pixel. The `time ordered' option was used, and a 25\,$\sigma$-threshold was applied. 
The level-2 products , i.e. fully reduced fits maps, were obtained using SCANAMORPHOS \citep{Roussel12} version 5.0 instead of applying the nominal HIPE highpass filtering and photProject reduction step, due to the improved performance of SCANAMORPHOS \citep[see ][]{Ragan12}.
The maps in each band were reduced with a pixel-size of $1\farcs0$/px.

\subsubsection{SPIRE}
The EPoS SPIRE scan maps at 250\,$\mu$m, 350\,$\mu$m and 500\,$\mu$m were obtained with a scan speed of 30''/s. They were reduced up to level-1 using HIPE version 5.0% (1982) 
with the calibration pool 5.1. The final level-2 reduction was done with SCANAMORPHOS version 9.0 using the `galactic' option. The pixel sizes are 6.0''/px at 250\,$\mu$m, 10.0''/px at 350\,$\mu$m and 14.0''/px at 500\,$\mu$m.

\subsubsection{Hi-GAL}
To include a larger coverage we obtained the raw data products from the Hi-GAL Open Time Key Project. These observations were carried out in the PACS/SPIRE parallel mode with a scan speed of 60$''$/s, resulting in large scale maps. The 70\,$\mu$m and 160\,$\mu$m PACS maps have been reduced using HIPE 8.0, applying the same options as for the EPoS PACS data. Level-2 processing has been performed with SCANAMORPHOS version 15. The pixel size is 3\farcs2/px for both maps. The 250\,$\mu$m, 350\,$\mu$m and 500\,$\mu$m SPIRE maps were reduced up to level-1 using HIPE version 8.0 with the calibration tree 8.1. SCANAMORPHOS version 15 was used to produce the level-2 maps. The pixel size is the same as for the EPoS SPIRE maps. The final maps are presented in the left panel of Figure \ref{fig:RGB_overview}. The Hi-GAL maps are used for the temperature mapping in Section \ref{sec:Tmap}.

\subsubsection{Photometry, point-source detection}
Due to the diffuse and heterogeneous background emission varying on a wide range of spatial scales, 
it is more robust to perform PSF photometry than aperture photometry.
Therefore the \texttt{starfinder} PSF photometry software \citep{Diolaiti00} was used for the point source detection and flux extraction for the PACS bands. 
The latest empirical PSF of the Vesta asteroid was used (OD 160), obtained from the Herschel Science Archive (HSA), since an empirical PSF derived from sources in \G48 is noise dominated and therefore less acceptable. For consistency the PSF maps were reprocessed with SCANMORPHOS and rotated to the orientation of the maps.
The noise in the PACS maps is not white noise, since it suffers from pixel-to-pixel correlations induced, e.g., by non-thermal low-frequency noise. 
To calculate the average noise in the map a Gaussian distributed noise was assumed, but corrected by a factor for the pixel-to-pixel correlation. 

The major source of confusion in the source detection is the complicated background. In particular the bright background towards the GMC returns numerous false detections. 
We only include regions of the PACS maps that correspond to regions in the ATLASGAL map (see Section \ref{sec:submm}) with fluxes above a $3\, \sigma$ level.
To further improve the detection efficiency an unsharp-mask-filter was applied in each PACS-band by subtracting a median smoothed map from the respective science-frame.
 With this procedure we avoid false detections, and miss three obvious sources for the automated detection run. 
These sources were then included in the catalog identified by eye as clear detections. With the compiled fixed source positions the final photometry was performed on the non-filtered science frames. 
To be considered a detection a source must meet the following conditions: it must have a source flux above the $5\sigma$ noise level and be detected in all PACS bands within the FWHM of the 160\,$\mu$m PACS PSF ($\sim 11\farcs2$). The results of the point source extraction are reported in Section \ref{sec:pointsrc}.
The given fluxes are already color-corrected for the source temperatures\footnote{\raggedright See:  \url{http://herschel.esac.esa.int/twiki/pub/Public/PacsCalibrationWeb/cc_report_v1.pdf}}.
Further details of the source extraction are given in the EPOS high-mass overview paper \citep{Ragan12}.

\subsection{Spitzer} \label{sec:Obs_Spitzer}
We obtained the  \textit{Spitzer} IRAC and MIPS maps from the Galactic Legacy Infrared Mid-Plane Survey Extraordinaire \citep[GLIMPSE,][]{Benjamin03} and the MIPS Inner Galactic Plane Survey \citep[MIPSGAL,][]{Carey09}. The 24\,$\mu$m MIPSGAL maps were processed with the MIPS Data Analysis Tool \citep[DAT v3.10; ][] {Gordon05} starting from the \texttt{raw} data products. 

We used the 8\,$\mu$m IRAC map to create an extinction map.
PSF photometry was performed on this map, using the \texttt{starfinder} package. 
An averaged, empirical PSF was used to remove all the point sources in this field. The residual map was used to compute the optical depths. One of the key problem in calculating an optical depth in IR data, is to determine the foreground and background intensities. Several different methods are described in the literature 
\citep[e.g.][]{Simon06b,Butler09,Ragan09,Butler12}.  However, to estimate the optical depth in one single cloud we used the method described in \citet{Vasyunina09}.

\begin{equation}
\tau=\ln \left( \frac{I_0-I_\mathrm{fg}}{I_\mathrm{IRDC}-I_\mathrm{fg}} \right)
\label{eq:IRAC_tau}
\end{equation}
where $I_{\rm IRDC}$  is the flux derived from the IRDC directly,
$I_0$ is measured of ``off-cloud'' in a nearby region meant to represent the characteristic unattenuated MIR emission.  $I_0$ (and $I_{\rm IRDC}$) contains contributions from both the foreground and background (i.e. $I_0 = I_{\rm bg} + I_{\rm fg}$), and only $I_{\rm bg}$ is needed to computed the optical depth. For $I_{\rm fg}$, we assume the zodiacal light emission provided in the header of the IRAC data product. This is a lower limit to the true foreground. Nevertheless, the uncertainties in the column densities are dominated by what is assumed about the extinction law (factor 2 to 3), whereas the $I_{\rm fg}$ contributes about an error less than 50\%.

The point-source fluxes from MIPS 24\,$\mu$m used in Section \ref{sec:pointsrc} were also obtained with \texttt{starfinder}. For MIPS PSF photometry we utilized a model PSF computed by the \texttt{STINYTIM}\footnote{\raggedright 
See: \url{http://irsa.ipac.caltech.edu/data/SPITZER/docs/dataanalysistools/tools/contributed/general/stinytim/}} program, based on \texttt{TINYTIM} for the HST \citep{Krist02}. 

The spectrum, shown in Figure \ref{fig:spec}, was obtained from the Infrared Spectrograph (IRS) on-board the \textit{Spitzer Space Telescope}. The observation (AOT: 12093696) was taken from the Spitzer Heritage Archive (SHA) as part of the ``Infrared Dark Clouds in the Inner Galaxy'' program. 
 We used a pipeline partly based on the SMART-package and the FEPS spectral extraction tools \citep{Bouwman06,Swain08,Pitann11}. The background subtracted on-source spectrum and the background spectra from the off-source nod-positions is shown in Figure \ref{fig:spec}. 
Only the long wavelength-order (LL) is reduced, since the slit for the short wavelength did not cover any point source in the IRDC. The LL slit did only cover about 60\% of the SMART-source PSF, which leads to a flux leakage in the continuum flux. But line shapes and line fluxes should be unchanged for the source spectrum, except for a lower S/N-ratio. 
A strong [SIII] emission appears over the whole IRS-slit, leading to an artifact in the source spectrum  due to an over-subtraction of the background. The source spectrum was corrected for this artifact using a polynomial background fit\citep[see Section 3.2.3 in ][]{Pitann11}. 

\subsection{Sub-Millimeter imaging} \label{sec:submm}

We utilized sub-millimeter data from the SCUBA Legacy Catalogue \citep{DiFrancesco08}. The 450 and 850\,$\mu$m observations were part of the fundamental catalog with a effective beamsize of $17\farcs3$ and $22\farcs9$, respectively. 
Given the better sensitivity of the survey, the 870\,$\mu$m observations from the  ``APEX Telescope Large Area Survey of the Galaxy'' \citep[ATLASGAL, ][]{Schuller09} were used in the temperature fitting (Section \ref{sec:Tmap}) and to determine column densities (Section \ref{sec:Colmap}). The corresponding beam size for the ATLASGAL data is around $19\farcs2$. The RMS noise level for ATLASGAL is $\sim 50\,\mathrm{mJy}$.

\subsection{Millimeter line observations}

We obtained molecular line data using the 22 meter MOPRA radio telescope and the MOPRA spectrometer (MOPS), operated by the Australian National Facility (ATNF). The observations were performed with the 3\,mm band receiver (90\,GHz) in the ``On-the-fly Mapping'' mode (OTF) with a beam size of $36''$  at 86\,GHz and $33''$  at 115\,GHz. Using MOPS in the ``Zoom'' setup allowed us to observe 13 lines simultaneously ($^{13}$CS, C$_2$H, CH$_3$CCH, CH$_3$CN, H$^{13}$CN, HCO$^+$, H$^{13}$CO$^+$, HCCCN, HCN, HNC, HNCO, N$_2$H$^+$, SiO).

Most of these lines are good tracers for dense gas \citep[see ][for a detailed description of each molecular species]{Vasyunina11}.
However, the RMS noise for all MOPRA maps is only slightly below 0.1K, that allows us to detect line emission above
3\,$\sigma$ only in the HCN, HNC, HCO$^+$ and N$_2$H$^+$ maps.

The spectrometer configuration results in a velocity resolution of $\sim 0.11$\,kms$^{-1}$. The main beam efficiency varies between  $49\%$ at 86\,GHz and $44\%$ at 100\,GHz \citep{Ladd05}. The typical system temperatures were between 210 and 290\,K. 
The $^{13}$CO~(1-0) data products were obtained from the Galactic Ring Survey \citep[GRS, ][]{Jackson06} performed by the Five College Radio Astronomy Observatory (FCRAO). With a angular resolution of $46''$ the beam of FCRAO is somewhat larger than that of MOPRA. The RMS noise level in the GRS map is below 0.9\,K. 

We obtained \ce{^{13}CO}~(2-1) and \ce{C^{18}O}~(2-1) maps using the 230\,GHz heterodyne receiver array \citep[HERA, ][]{Schuster04} at the IRAM 30\,m telescope. The average FWHM of the HERA beam is 11\farcs3, which is much better compared to MOPRA or FCRAO. The RMS noise level is below 0.5 and 0.3\,K  in the \ce{^{13}CO} and \ce{C^{18}O} HERA maps, respectively. 

To acquire methylacetylene (\ce{CH3C#CH}) data observations were performed with the IRAM 30\,m telescope in position switching mode.
For these observations the EMIR receiver with Fast Fourier Transform Spectrometer (FTS) as a backend was used.
Total time On source and Off source was 33\,min.
The Off source position was chosen 600$''$ away from the On source position.
The On source position was pointed at $19^\mathrm{h}21^\mathrm{m}49^\mathrm{s}.9$  $+13^\circ 49'34\farcs7$ (FK5, J2000). 
The pointing was checked every hour, giving a pointing accuracy better than 3$''$.
The typical system temperature during observations was about 200\,K.
At 153\,GHz the telescope beam size is 16$''$ and the beam efficiency is 0.73.
The data were analyzed using GILDAS/CLASS software.

A list of physical parameters for each observed line is given in Table \ref{tab:line_prop}.

\section{Results} \label{sec:Results}

\begin{deluxetable*}{lcccccclll}%[th]
\tablecolumns{10}
\small
\tabletypesize{\scriptsize}
\tablewidth{0pt}
\tablecaption{Molecular lines}
\tablehead{
\colhead{Molecule}	& \colhead{Trans.}	& \colhead{$\nu_{r}$}	& \colhead{$\Delta V_{therm}$}	& \colhead{$A_{ul}$}		& \colhead{$E_u/k_B$} 	& \colhead{$n_{crit}$} & \colhead{beam size} & \colhead{Instrument} & \colhead{Obs. Date} \\
					&					& \colhead{GHz}   		& \colhead{kms$^{-1}$}	& \colhead{$10^{-5}$\,s$^{-1}$}  & \colhead{K}		& \colhead{cm$^{-3}$}  & \colhead{arcsec} &	& }
\startdata 
HCO$^+$ 			& $1-0$	&	89.188523	& 0.166		& 3.0		& 4.28		& $2.1\cdot 10^5$ 	& $38''$ 	& MOPRA &  2010 Jul 7-9 \\
HCN					& $1-0$	&	88.631602	& 0.172		& 2.4		& 4.25		& $1.5\cdot 10^6$ 	& $38''$ 	& MOPRA &  2010 Jul 7-9  \\
HNC					& $1-0$	&	90.663568	& 0.172		& 2.690		& 4.35 		& $3.2\cdot 10^5$	& $38''$ 	& MOPRA &  2010 Jul 7-9  \\
N$_2$H$^+$			& $1-0$	&	93.173700	& 0.166		& 3.6		& 4.47		& $1.8\cdot 10^5$ 	& $38''$ 	& MOPRA &  2010 Jul 7-9  \\
$^{13}$CO			& $1-0$	&   110.201354 	& 0.166		& 0.0063	& 5.29		& $1.9\cdot 10^3$	& $44''$	& FCRAO & Archive data\,$^1$ \\
$^{13}$CO			& $2-1$	&   220.398684 	& 0.166		& 0.0604 	& 15.87		& $2.0\cdot 10^4$	& $11\farcs25$	& IRAM/HERA & 2012 Jan 23 \\
C$^{18}$O			& $2-1$	&   219.560319 	& 0.163 	& 0.0601	& 15.81		& $1.9\cdot 10^4$	& $11\farcs25$	& IRAM/HERA & 2012 Jan 23 \\
\ce{CH3C2H}			& 9$_0-8_0$	&	153.8172			& 0.142		& 8.75		&			&					&		$16''$				& IRAM/EMIR-FTS	& 2011 Jun 10/12	\\
\ce{CH3C2H}			& 9$_1-8_1$	&	153.8142			& 0.142		& 8.44			&			&					&		$16''$				& IRAM/EMIR-FTS	& 2011 Jun 10/12	\\
\ce{CH3C2H}			& 9$_2-8_2$	&	153.8054			& 0.142		& 7.55			&			&					&		$16''$				& IRAM/EMIR-FTS	& 2011 Jun 10/12	\\
\ce{CH3C2H}			& 9$_3-8_3$	&	153.7907 			& 0.142		& 12.52			&			&					&		$16''$				& IRAM/EMIR-FTS	& 2011 Jun 10/12	\\
\hline
CH$_3$C$_2$H		& $5_1-4_1$			&	85.455665	& 0.142		& 0.1948	& 19.53		&					& $38''$ 	& MOPRA &  2010 Jul 7-9  \\
H$^{13}$CN			& $1-0$				&	86.339860	& 0.169		& 2.2256	& 4.14		& $2.6\cdot 10^6$	& $38''$ 	& MOPRA &  2010 Jul 7-9  \\
H$^{13}$CO$^+$		& $1-0$				&	86.754288	& 0.164		& 3.8534	& 4.16		& $1.9\cdot 10^5$	& $38''$ 	& MOPRA &  2010 Jul 7-9  \\
SiO					& $2-1$				&	86.846960	& 0.135		& 2.927		& 6.25		& $1.9\cdot 10^6$	& $38''$ 	& MOPRA &  2010 Jul 7-9  \\
C$_2$H				& $2-1$				&	87.316925	& 0.179		& 0.1528	& 4.19		& $2\cdot 10^5\,^\mathrm{(a,b)}$		& $38''$ 	& MOPRA &  2010 Jul 7-9  \\
HNCO				& $4_{0,4}-3_{0,3}$	&	87.925238	& 0.137		& 0.9024 	& 10.55		& $4.5\cdot 10^6$	& $38''$ 	& MOPRA &  2010 Jul 7-9  \\
HC$_3$N				& $10-9$			&	90.978865	& 0.125		& 5.812		& 24.01		& $7.5\cdot 10^4$	& $38''$ 	& MOPRA &  2010 Jul 7-9  \\
CH$_3$CN			& $5_1-4_1$			&	91.985316	& 0.140		& 6.0800	& 20.39		& $4\cdot 10^5\,^\mathrm{(b)}$	& $38''$ 	& MOPRA &  2010 Jul 7-9  \\
$^ {13}$CS			& $2-1$				&	92.494308	& 0.134		& 1.4124	& 6.66		& $5.0\cdot 10^5$	& $38''$ 		& MOPRA  &  2010 Jul 7-9  
\enddata 
\tablecomments{The columns list, the molecular species, the transition, the rest frame frequency, the thermal line width at 17.3K, the Einstein coefficient, the upperstate energy, the critical densities and the angular resolution for our observations, the used instrument, from left to right, respectively. The upper part of the table presents the detected molecular lines, while the bottom part lists the MOPRA lines lacking a $3\,\sigma$-detection.} \tablenotetext{1}{from the  Galactic Ring Survey with FCRAO}
%\tablenotetext{2}{from HERA/IRAM}
\tablenotetext{a}{from \citet{Lo09}}
\tablenotetext{b}{from \citet{Sanhueza12}}
\label{tab:line_prop}
\end{deluxetable*}

\begin{figure*}
	\centering
	\includegraphics[width=0.85\textwidth, angle=270]{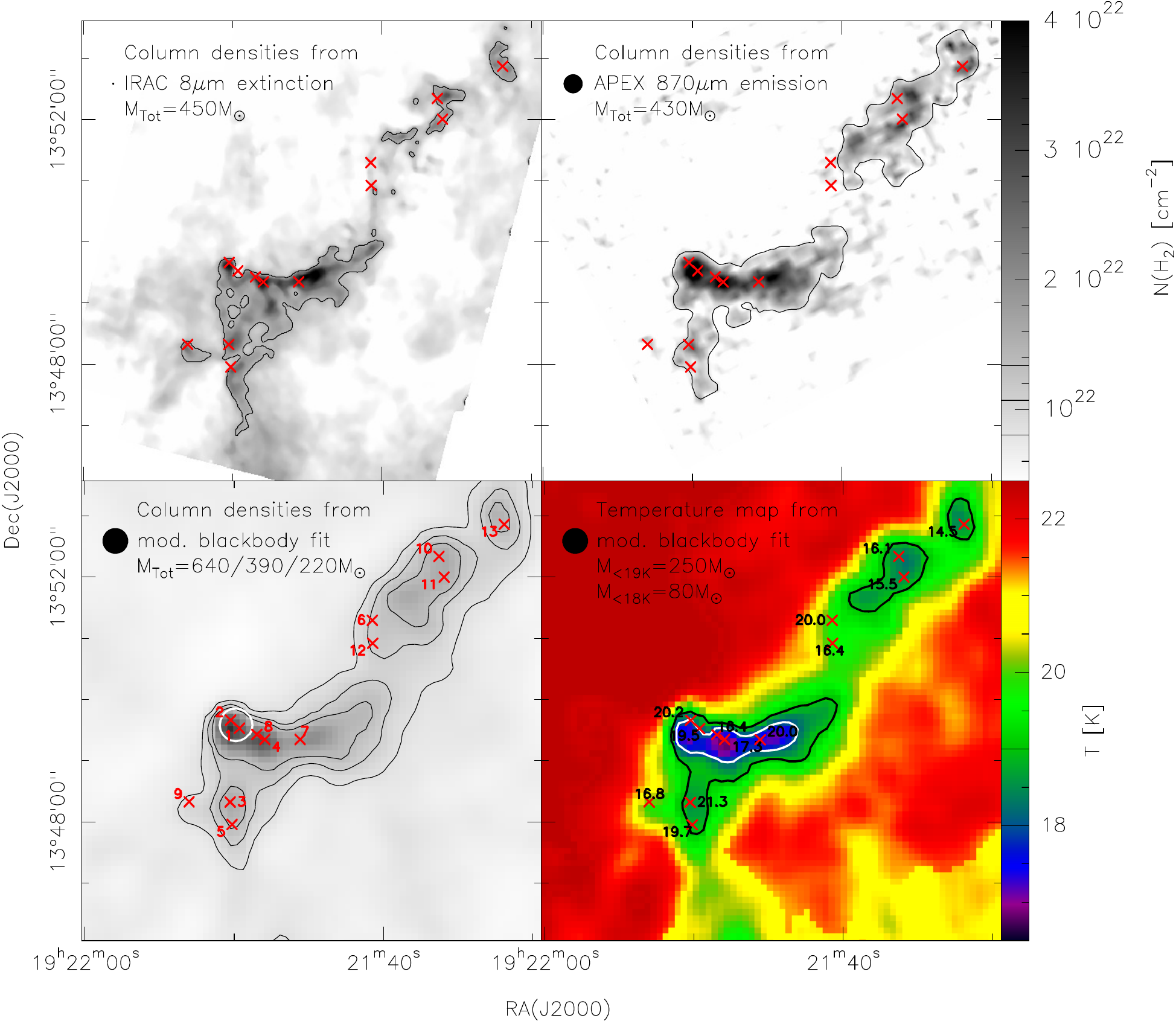}
	\caption{The top two panels show \ce{H_2} column density maps obtained from IRAC 8\,$\mu$m extinction (left) and ATLASGAL 870\,$\mu$m (right) observations. 
 In these top panels the column density thresholds for the mass determination are shown as contours: within the contour of ${1.34\cdot 10^{22}\,\textrm{cm}^{-2}}$ a mass of 450\,M\sun ~is enclosed for IRAC 8\,$\mu$m while the ${0.85\cdot 10^{22}\,\textrm{cm}^{-2}}$ ATLASGAL contours contain 430\,M$_\odot$ of material.
 The bottom two panels show the column density (left) and temperature (right) maps from the modified blackbody fitting (Section \ref{sec:Tmap} and \ref{sec:Colmap}) of the \textit{Herschel}, SCUBA and ATLASGal maps. 
 The point sources from Section \ref{sec:pointsrc} are indicated by \textcolor{red}{$\times$}. The contours indicate column densities of ${0.85\cdot 10^{22}}$, ${1.10\cdot 10^{22}}$, and ${1.34\cdot 10^{22}\,\textrm{cm}^{-2}}$ and encircle 640, 390, and 220\,M$_\odot$ respectively. The white circle indicates the pointing and beam size of the \ce{CH3C2H} observations, which indicates a gas temperature of $26\pm7$\,K.
	The temperatures from the point source fitting are indicated in the last plot. This temperature map is overplotted with isothermal contours of 19\,K (black) and 18\,K (white). These isotherms incorporate 250\,M$_\odot$ and 80\,M$_\odot$ for 19\,K and 18\,K, respectively.}
	\label{fig:NH2_Tmap}
\end{figure*}

\subsection{Morphology - From large to small scales} \label{sec:morpho}
The left panel of Figure \ref{fig:RGB_overview} shows the galactic neighborhood of \G48. The image is composed of the 70, 160 and 250\,$\mu$m Hi-GAL maps in blue, green and red, respectively. A global brightness gradient towards the outer parts of the galactic plane is present over the whole image.
In the vicinity of \G48 the GMC W51 is observed. The cloud complex W51A and W51B have an angular distance of $\sim 15'$ and $\sim 50'$ to \G48, respectively. The EPoS maps (Figure \ref{fig:RGB_overview}, top right panel), in particular the PACS observations, show a brightness gradient towards W51 and the infrared bubble N100 \citep{Churchwell06}. W51 and \G48 are not physically connected since recent observations place W51 twice as far away as \G48 (see next Section) 

In the \textit{Spitzer} IRAC 8\,$\mu$m and the MIPS 24\,$\mu$m channel the IRDC appears as the typical dark filamentary absorption feature against the diffuse galactic emission background (see Figure \ref{fig:stamps}). \G48 is extended over $\sim 7\farcm5$ in the north-south direction. The dark cloud can be roughly divided into three parts. The central region of the IRDC appears as a elongated structure of $\sim 2\farcm3$ in east-western orientation. This region has the brightest sub-millimeter emission.
The western part of the central region shows the highest extinction in the \textit{Spitzer} maps. Only this part of the IRDC appears in clear absorption in all three PACS channels. 
Furthermore this part is clearly associated with the most massive sub-millimeter clump, called ``extinction peak'' (EP, with a mass of $\sim 130$\,M$_\odot$ from SCUBA observation) in \citet{Ormel05}.
The eastern part of the central IRDC only shows clear extinction in the \textit{Spitzer} bands. In the \textit{Herschel} PACS maps this region is dominated by compact background emission from embedded point sources, in particular Source~1 (see Section \ref{sec:pointsrc}). The central region, as the rest of \G48, appears as a diffuse emission structure in the SPIRE maps, but two compact sources can be identified in the eastern central part. 
The peak positions of these emission features detected at 250$\mu$m and 350$\mu$m  are in good agreement with the peak positions of two sub-millimeter clumps (P1 with 91 M$_\odot$ and P2 with 100 M$_\odot$) identified by \citet{Ormel05}.
Due to the lower resolution of the 500\,$\mu$m SPIRE channel the two positions are not resolved.

The structures found south- and northwards of the central part of the IRDC appear more diffuse in the \textit{Spitzer} bands and are barely recognizable in the PACS channels. In contrast, in the SPIRE bands this diffuse absorption features appear as an elongated emission structure. While this emission is diffuse in the northern part, compact emission is detected in good agreement with the SCUBA and ATLASGAL data in the southern part of the IRDC. The associated sub-millimeter peak was called ``P3'' in \citet{Wiel08} (see Figure \ref{fig:stamps}).

\subsection{Distance}  \label{sec:dist}
The LSR velocities found for the different clumps and molecular species do not differ by more than $3\,\sigma_V$. This results in a weighted mean velocity of $\langle V_{LSR}\rangle=33.91\pm 0.05$\,kms$^{-1}$ (see Section \ref{sec:molecline}). 
The resulting kinematic distance for \G48 is  $2.6\pm 0.6$\,kpc. This value is used for all calculations in this paper. 
We derive kinematic distances using the \citet{Reid09} algorithm, which takes into account trigonometric parallax measurements of masers in high-mass star forming regions in the Galaxy, applied to a disk Galaxy model. We use their standard derived rotational parameters of $\Theta_0$ = 254\,km s$^{-1}$ and $R_0$ = 8.4\,kpc.

The Hi-GAL three color image (Figure \ref{fig:RGB_overview}) shows
\G48 at a close projected distance to the GMC W51 and several objects, as e.g., HII regions, infrared bubble and other bright point-sources.
Water maser observation estimates a distance of $\sim$5.4\,kpc for W51 \citep{Sato10}.
Known distances of infrared bubbles (Sarah Kendrew 2012, private communication) and sources from the RMS survey catalog \citep[e.g., massive YSO, HII regions ][]{Urquhart08} in the vicinity of \G48 have a distance of more than $5$\,kpc. 
Moreover, no CO is detected in the global GRS maps outside of \G48 at its typical LSR velocities.

We therefore conclude that \G48 is an isolated region making it an unique IRDC.

\subsection{Herschel temperature map} \label{sec:Tmap}

To create temperature and column density maps of \G48 we used \textit{Herschel} maps and ground based sub-millimeter data. 
In brief, the basic steps we have taken to fit the temperature and column density maps are as follows:
\begin{enumerate}
\item The global background flux offsets are estimated from the large scale Hi-GAL maps and subtracted from the maps.
\item All data (PACS at 160\,$\mu$m, SPIRE at 250 and 350\,$\mu$m, SCUBA at 450\,$\mu$m, and ATLASGAL 870\,$\mu$m) were convolved to a common limiting beam size (SPIRE 350\,$\mu$m) and converted to uniform units (Jy/arcsec$^2$). The inclusion of the also available SPIRE 500\,$\mu$m data would force us to commonly convolve all other maps to it coarse resolution of around $36''$. We therefore neglect these 500\,$\mu$m data for the temperature mapping.
\item In the final fitting process an SED was fitted to each pixel using a modified black body function, requiring at minimum three data points for the SED.
\end{enumerate}

For the final pixel-to-pixel SED fitting a careful preparation of the initial data is needed. A complete description of the data treatment and fitting process is given in \citet{Launhardt12}. 
The observed globules in \citet{Launhardt12} are more isolated and suitable background flux offsets can be calculated within the EPOS maps. Unfortunately, the larger extent of the IRDC, the imprint of the adjacent GMC and infra-red bubble (see Section \ref{sec:morpho}), and the limited map sizes prevent a reasonable background determination. To overcome these limitations we utilized \textit{Herschel} maps from the Hi-GAL program which cover a much larger area.
We extracted the background fluxes from a low-emission region off the galactic plane located at $19^\mathrm{h}23^\mathrm{m}26^\mathrm{s}.018$ $+12^\circ37'28\farcs06$ (J2000) with a box size of 500$''\,$. 
The actual method to extract the background fluxes in this field with relatively low spatial variations is adopted from \citet{Stutz09}. It is an iterative Gaussian fitting algorithm applied to the pixel flux distribution in the selected region. The best fit to the background levels is found by omitting the non-Gaussian components.
Finally background levels of 0.040\,Jy/pixel and 1.434, 1.191\,Jy/beam were computed for 160, 250 and 350\,$\mu$m, respectively. 

\citet{Battersby11} used a large scale Gaussian profile to remove the Galactic background from their observed Hi-GAL maps toward two different fields before the SED fitting. \textit{Herschel} temperature maps of a sub-region close to the star-forming complex W43 were created by \citet{Beuther12} without any global background subtraction. For lower temperatures ($< 25$\,K) both methods just differ by $1-2$\,K. % (Beuther, priv. communication). 
Our maps appear to be more sensitive to the background determination\footnote{We also determined the background offsets from a field with low spatial variations and almost no compact emissions inside the galactic plane. But the resulting temperature maps from the SED fitting were suspiciously high ($>20$\,K inside the IRDC). Therefore, we decided to estimate  the background offsets further outside of the galactic plane.}, however changing our estimated background level by $\pm 50\%$ does not have a significant impact on the derived temperatures.

For the SED fitting all  \textit{Herschel} maps (PACS~160\,$\mu$m, SPIRE~250 and 350\,$\mu$m) were convolved to the SPIRE~350\,$\mu$m beam, using the kernels from \citet{Aniano11}. % \citet{Gordon08}. 
For the sub-millimeter maps (SCUBA 450\,$\mu$m and APEX~870\,$\mu$m) a Gaussian kernel was used. For this step all maps were calibrated in units of Jy/arcsec$^2$.

In the resulting maps the remaining emission per pixel is assumed to be well represented by modified blackbody emission:
\begin{equation}
S_\nu(\nu)=\Omega\bigl(1-e^{-\tau(\nu)}\bigr)B_\nu(\nu,T_\mathrm{d}) 
%  -I_\mathrm{bg}(\nu)\bigr)
\label{eq:BBtmap}
\end{equation}
with 
\begin{equation}
\tau(\nu)=N_\mathrm{H}\, m_\mathrm{H}\, R\, \kappa(\nu)
\end{equation}
where $S_\nu(\nu)$ is the observed flux density for a given frequency $\nu$ and solid angle $\Omega$ (all fluxes are normalized to ${\mathrm{Jy/arcsec}^2}$), $\tau(\nu)$ is the optical depth, $B_\nu(\nu,T_\mathrm{d})$  the Planck function, $T_\mathrm{d}$ the dust temperature, %$I_\mathrm{bg}(\nu)$ the background flux,
$m_\mathrm{H}$ the proton mass, $R$  the dust-to-gas mass ratio, and ${N_H=2\cdot N(\mathrm{H_2})+N(\mathrm{H})}$ the total hydrogen column density. For the column density map (Figure \ref{fig:NH2_Tmap}, bottom left) we assume that all hydrogen is present as molecular gas in \G48. For the dust opacity $\kappa$ we adopt the model from \citet{Ossenkopf94} for thin ice mantels and a gas density of $10^5$\,cm$^{-3}$.
For the hydrogen-to-dust mass ratio we adopted a value of $R=110$ \citep[e.g., ][]{Sodroski97}. To account for helium and heavy elements one would choose the total gas-to-dust mass ratio 1.36 times higher.
A modified black body function, as in Equation \ref{eq:BBtmap} was fitted individually for each pixel using $N_\mathrm{H}$ and $T_\mathrm{d}$ as free parameters with a $\chi^2$-minimization. 
The Hi-GAL data includes only the 70 and 160\,$\mu$m maps for the PACS bands. To avoid contributions from stochastically heated small grains or optical thick emission, we exclude the 70\,$\mu$m band \citep[e.g.,][]{Battersby11,Pavlyuchenkov12}. 
Usually the fluxes in the three remaining \textit{Herschel} bands (160, 250, and 350\,$\mu$m) are well constrained and provide a good fit to the modified black body function. 
Therefore, every shown pixel in the temperature map (Figure \ref{fig:NH2_Tmap}) results from an SED fit of at least these three Herschel bands. If one of the submillimeter (450 or $870\,\mu$m) fluxes deviate by more than $3\,\sigma$ from the SED-fluxes constrained in the other bands it is not used in the temperature and column density calculation.

The temperature map (Figure \ref{fig:NH2_Tmap}, bottom right) shows a steep drop at the boundaries of \G48 within the resolution limits set by the size of the convolution kernels (a 19\,K contour is overplotted on the IRAC map at 8\,$\mu$m and PACS at 100\,$\mu$m in Figure \ref{fig:stamps}). In the central region of \G48, with the strongest sub-millimeter emission, the temperatures drop below 18\,K (white contour in Figure \ref{fig:NH2_Tmap}, bottom right). The local minima agree with the position of the sub-millimeter peaks P1 (17.1\,K), P2 (16.8\,K), and EP (17.1\,K). 
In general our observed line-of-sight temperatures are higher than the ones (10-15\,K) found by \citet{Ormel05} for P1, P2, and EP. However, the later temperature profiles from \citet{Ormel05} are obtained from fitting observational results to core models \citep[similar to][see Section \ref{sec:temp_disc}]{Wilcock12}.

To assess the uncertainties of our line-of-sight-averaged temperatures, we investigated how the change of different parameters influence the resulting temperatures and column densities. When altering the subtracted global background level, that is subtracted from the Herschel maps, by $\pm 50\%$\ before the fitting process and using different beam sizes for the convolution, the resulting temperatures are found to differ by no more than 0.3\,K inside the IRDC, where dust emission is detected in the ATLASGAL maps.
In the temperature fitting routine, we assume mean absolute background flux levels that were derived for regions well outside the galactic plane \citep[see ][]{Launhardt12} and might not be appropriate for this IRDC. For lack of better estimates of the local absolute background levels, we have tested the effect of simply amplifying those background levels by a factor of 100. The resulting temperatures and column densities in the IRDC were found to change by no more than 0.1\,K and 2.5\%, respectively. The effect would be stronger for lower column densities, which we do not consider here. Hence, we can robustly conclude that the effect of the uncertain knowledge in the absolute local background flux levels on the derived temperatures and column densities in the IRDC is smaller than the uncertainties introduced by the calibration, data reduction, and dust opacity models, and is thus negligible.

The total relative uncertainties introduced by calibration and data reduction are found to be 2.5\% for the temperatures and 10\% for the column densities. The largest uncertainties are introduced by the uncertain knowledge of the underlying dust opacity model. Therefore, we assume an overall temperature uncertainty of $\pm 1$\,K. A more detailed discussion of the different contributions to the uncertainties can be found in \citet{Launhardt12}.

It should be mentioned that the temperatures outside \G48 are not very reliable because of the brightness gradient introduced by W51 and N100, as mentioned earlier. In the column density and temperature maps (Figure \ref{fig:NH2_Tmap}) the point source positions and temperatures are indicated. 

The temperatures derived at the locations of the point sources in the large--scale temperature map are discrepant from the temperatures derived by fitting the individual SEDs of said point-sources, the later of which were extracted at native resolution. This discrepancy is not surprising as it is caused by beam convolution in the temperature map and the inclusion of longer wavelengths.

\subsection{Column densities} \label{sec:Colmap}
To estimate the column densities for \G48 we utilized three different methods; mid-infrared extinction map from IRAC at 8\,$\mu$m,  sub-millimeter dust emission and modified blackbody fitting including \textit{Herschel} and sub-millimeter data (see Section \ref{sec:Tmap}).

The IRAC 8\,$\mu$m extinction maps provides us with an estimate of the optical depth per pixel (see Section \ref{sec:Obs_Spitzer}). These optical depths were converted into column densities (Figure \ref{fig:NH2_Tmap}, bottom left) using \citep{Vasyunina09}:
\begin{equation}
N_\mathrm{H_2}=1.086\;\tau/C_{ext}
\label{eq:coldens_tau}
\end{equation}
Where $\tau$ is the optical depth per pixel and $C_{ext}$ is the extinction cross-section.
The nature of the mid-infrared extinction law below 10\,$\mu$m in IRDCs and massive star-forming regions is still a matter of debate \citep[e.g., ][]{Butler09}. In agreement with findings of \citet{Indebetouw05} we used an extinction factor of $R_V=5.5$ to parametrize the extinction cross-section $C_{ext}$. We choose ${C_{ext}=4.62\cdot 10^{-23}\,\mathrm{cm}^2}$ from \citet[][(case B)]{Weingartner01}. The derived column densities also depend on the assumed $R_V$ value and, therefore, the used extinction cross-section. The extinction factor $R_V=3.1$ is more representative of the diffuse ISM and would result in significantly higher column densities. Another effect for the 8\,$\mu$m extinction is the saturation of  the column density for high values, therefore the masses based on IRAC~8\,$\mu$m discussed further below can only be lower limits \citep[for a detailed discussion, see][]{Vasyunina09}.

The second method to determine the hydrogen column densities utilizes the dust emission detected in the 870\,$\mu$m map from ATLASGAL:
\begin{equation}
N_\mathrm{H_2}=\frac{RF_\nu}{B_\nu(\nu,T_d)m_\mathrm{H_2}\kappa\,\Omega}
\label{eq:coldens_submm}
\end{equation}
where $F_\nu$ is the sub-millimeter flux, $R=110$ the gas-to-dust ratio, $B_\nu(\nu,T_d)$ the Planck function for a given wavelength (870\,$\mu$m) and dust temperature ($T_d=17.3$\,K or 19.2\,K, see below),  $m_\mathrm{H_2}$ the mass per hydrogen molecule, and $\Omega$ the beam size. We adopt a mass absorption coefficient of $\kappa=1.47\,$cm$^2$g$^{-1}$ from \citet[thin ice mantel, $10^5\,$cm$^{-3}$, as in Section \ref{sec:Tmap}]{Ossenkopf94}.

We have three independently derived column density maps: IRAC~8\,$\mu$m, ATLASGAL 870\,$\mu$m, and the SED fitted map combining \textit{Herschel} and ATLASGAL.  Each one requires
us to make different and independent assumptions in order to derive column
densities; furthermore, they also vary widely in angular resolution. 

For example, the 8\,$\mu$m extinction map, which has a resolution of $\sim 3''$, requires an
assumption of a certain extinction factor for the dust extinction cross-section, optically thin emission and the local background contribution ($I_\mathrm{BG}$ in Equation \ref{eq:IRAC_tau}), but is temperature independent. 
Furthermore, the residuals of the PSF removal for bright IRAC point-sources with compact background emission can result in an underestimate of local column density (see e.g., source 1 in Figure \ref{fig:NH2_Tmap}, top left).

For the ATLASGAL 870\,$\mu$m map, with a resolution of $\sim 19$'', always a fixed temperature is assumed, it requires optically thin emission and a dust opacity  choice at the wavelength of interest.

While the SED-fitted map also requires a dust model assumption, it contains information from multiple wavelengths and simultaneously derives $N_\mathrm{H_2}$ and temperature for each pixel, under the assumption that the mass is well-represented by a modified black body and assuming a fixed spectral index $\beta$. Therefore the derived column densities can be considered more robust compared to the two previous methods.
However, the SED-fitted map also has the lowest resolution of the three methods considered here, with final convolved beam size of 25$''$, equivalent to $\sim0.3$\,pc at the distance of \G48.  Here we discuss the integrated masses for these three independent methods.

The total mass of \G48 can be calculated within a contour of interest using the column densities mask (shown in Figure \ref{fig:NH2_Tmap}):
\begin{equation}
M=\sum_{i} a_\mathrm{px}\;m_{\mathrm{H}_2}\;N_{i}(\mathrm{H}_2)
\label{eq:mass_submm}
\end{equation}
where $a_{px}$ is the pixel area, $m_{\mathrm{H}_2}$ and $N_{i}(\mathrm{H}_2)$ the column density per pixel.

The masking contours for the IRAC $8\,\mu$m extinction map were chosen by visual inspection. This contour results in a threshold of ${32.15\,\mathrm{MJy/sr}}$, which correspond to a column density of ${1.34\cdot 10^{22}\,\mathrm{cm}^{-2}}$.
A total \ce{H2} mass of $450$\,M$_\odot$ is enclosed in these contours (see Figure \ref{fig:NH2_Tmap}, top left panel).

We assume a temperature of 17.3\,K (the median point source
temperature, see Section \ref{sec:pointsrc}) for the ATLASGAL derived
column density map (Figure \ref{fig:NH2_Tmap}; top right panel).  The
column density threshold is adopted to be at 3$\times$the RMS noise
level ($\sim 50$\,mJy/beam), corresponding to ${N(\mathrm{H_2}) = 0.85\cdot
10^{22}\,\mathrm{cm}^{-2}}$.  The resulting total mass inside this column
density contour is ${M(\mathrm{H}_2) = 430\,\mathrm{M}_\odot}$.  For comparison,
the median {\it Herschel} derived temperature (Figure
\ref{fig:NH2_Tmap};bottom right panel) in this same region is 19.2\,K.
If we adopt this temperature instead of 17.3\,K, as assumed above, the
column density threshold becomes ${N(\mathrm{H}_2) = 0.75\cdot
10^{22}\,\mathrm{cm}^{-2}}$ and the total enclosed ATLASGAL mass is ${M(\mathrm{H}_2) =
675\,\mathrm{M}_\odot}$.  

For reference, applying a column density threshold of
${1.34\cdot10^{22}\,\mathrm{cm}^{-2}}$ (derived from the from IRAC 8\,$\mu$m
column density map) is equivalent to applying a 18\,K isotherm to the
{\it Herschel} derived column density map, and results in thresholds
that encompass approximately the same spatial extent of the three
SCUBA peaks from \citet{Ormel05}. The mass enclosed within these
thresholds differ by only $20-30\%$ compared to the total mass
(320\,M\sun) of these three peaks.  

The total mass for \G48 derived from the 8\,$\mu$m extinction
(${450\,\mathrm{M}_\odot}$) and the 870\,$\mu$m (${430\,\mathrm{M}_\odot}$) emission are
remarkably similar. Furthermore, the average column density threshold
for both maps corresponds to a \textit{Herschel} derived total mass of ${M(\mathrm{H}_2)
=390\,\mathrm{M}_\odot}$, also in excellent agreement with the other
estimates. Considering the uncertainties in the various assumptions,
such as the dust properties, uncertainties in background flux levels,
and variations in beam sizes we find that our mass agreement for \G48
is strikingly good across different methods and compared to previous
work.

In general the column density for the diffuse emissions in \G48 are
comparable for all three methods. For EP, P1, P2, P3 and P4 the peak
column densities agree within 18\% for the ATLASGAL and the \textit{Herschel}
maps. However, the peak column densities derived from the 8\,$\mu$m
extinction are significantly higher then the fitted \textit{Herschel} maps. The
peak columns can be higher up to $75\%$. This might be an effect introduced by the higher spatial resolution of IRAC. 

\subsection{Point source Classification} \label{sec:pointsrc}

\begin{figure*}
	\centering
	\plotone{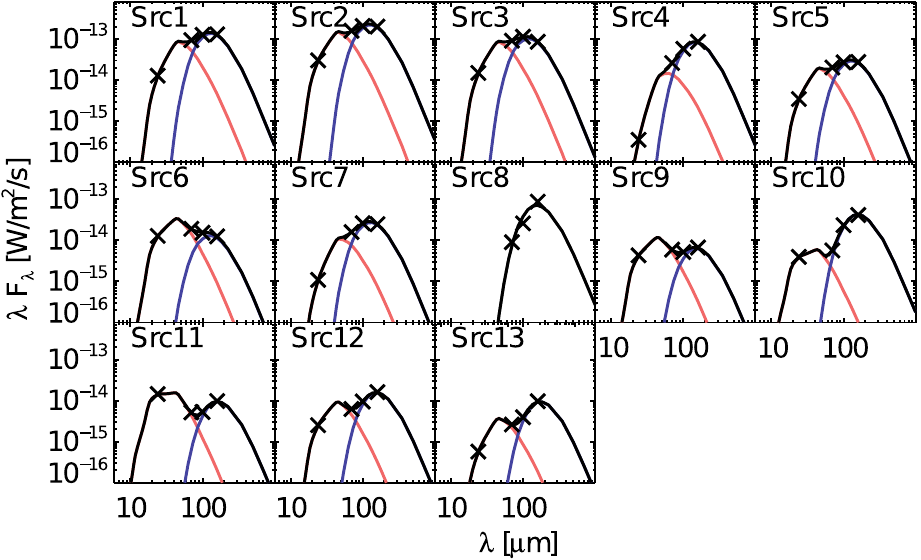}
	\caption{The modified blackbody fits for the detected sources (see Section \ref{sec:pointsrc}).}
	\label{fig:blackbodies}
\end{figure*}

\begin{deluxetable*}{ccc|rrrr|rrrr|r}[th]
\tablecolumns{12}
\small
\tabletypesize{\scriptsize}
\tablecaption{Point sources}
\tablehead{
\colhead{Source} & \colhead{R.A. (2000)} & \colhead{Dec (2000)} & \colhead{$F_{24\mu m}$}  & \colhead{$F_{70\mu m}$ } & \colhead{$F_ {100\mu m}$} &\colhead{$F_{160\mu m}$} & \colhead{$T_L$} & \colhead{$L_L$}     & \colhead{$M_L$}  &  \colhead{$L_H$} 		& \colhead{Comments$^\mathrm{(a)}$} \\ 
				 & \colhead{[hh:mm:ss[}  & \colhead{[dd:mm:ss]} & \colhead{mJy}  		   & \colhead{mJy} 		      & \colhead{mJy} 			  & \colhead{mJy} 			& \colhead{K}     & \colhead{L$_\odot$} & \colhead{M$_\odot$} & \colhead{L$_\odot$}	& \\ }
\startdata
1   & 19:21:49.65 & +13:49:31.26   &  102.7  & 1764    & 4086    & 6931  & 18.3 & 48.2 & 14.3 & 36.1 &  S5, P1 \\ 
2   & 19:21:50.24 & +13:49:39.28   &  241.1  & 3702    & 6931    & 10712 & 17.9 & 35.5 & 11.9 & 38.0  &  S6, P1 \\
3   & 19:21:50.26 & +13:48:18.96   &  119.5  & 1730    & 3659    & 4745  & 20.9 & 17.7 &  2.4 & 14.6 &  S2, P3 \\ 
4   & 19:21:47.95 & +13:49:20.37   &    2.8  & 460     & 1807    & 4717  & 17.2 & 15.1 &  6.3 & 2.8  & S15, P2 \\ 
5   & 19:21:50.13 & +13:47:57.00   &   27.7  & 474     & 902     & 1474  & 19.8 &  5.5 &  1.0 & 4.5  &  S1  	\\
6   & 19:21:40.76 & +13:51:17.42   &  102.7  & 365     & 494 	 & 693   &  16.8 & 2.4 &  1.2 & 6.5  &     	\\ 
7   & 19:21:45.58 & +13:49:20.51   &    8.7  & 299     & 842 	 & 1370  & 19.5 &  4.8 &  1.0 & 1.7  & S12, EP \\ 
8   & 19:21:48.48 & +13:49:25.26   &      -  & 210     & 865     & 4668  & 17.4 & 13.3 &  5.3 &   -  & P2    	\\
9   & 19:21:53.01 & +13:48:19.15   &   34.7  & 139     & 174     & 357   & 16.5 &  1.1 &  0.6 & 2.8  &  S3  	\\ 
10  & 19:21:36.30 & +13:52:20.23   &   31.8  & 132     & 780     & 2209  & 16.8 &  7.9 &  3.8 & 1.6 & S11 	\\ 
11  & 19:21:35.95 & +13:51:59.97   &  115.5  & 121     & 177     & 521   & 15.6 &  1.7 &  1.3 & 4.4 &  S9  	\\ 
12  & 19:21:40.73 & +13:50:54.91   &   20.2  & 115     & 303 	 & 894   & 15.6 &  3.2 &  2.4 & 1.5  & S13 	\\ 
13  & 19:21:31.93 & +13:52:51.82   &    4.6  & 61      & 131     & 516   & 14.5 &  1.7 &  1.9 & 1.0  &     	
\enddata
\tablecomments{$^\mathrm{(a)}$~Sx are the associated MIPS 24\,$\mu$m from \citet{Wiel08}, Px and EP are the associated sub-millimeter peaks.}
\label{tab:pnt_src}
\end{deluxetable*}

In Table \ref{tab:pnt_src} we present the fluxes of the 13 point sources which were detected in all three PACS bands. Except for source 8, all PACS detections have counterparts in MIPS 24\,$\mu$m. Source 13 was not included in the \citet{Wiel08} sample, but is detected as a point source at 24\,$\mu$m (E2 in Figure \ref{fig:stamps}).  If one allowed point source detection in only two PACS bands this would add only one more obvious point source. 
This source is also detected at 24 micron (E1) but appears as a diffuse blob of emission at 100 micron, not well-matched to the characteristic PSF we expect. 
However, the projected distance to Source 13 (E2) is $\sim 0.24$\,pc which is close to the median separation discussed in Section \ref{sec:conclusion}, and therefore would not change the outcome of this discussion. 
All detected sources are aligned along the mid-infrared filaments of the IRDC. Even though the emission gradient towards the GMC (on the eastern edge of our maps) is brighter and contains some compact emission features, no PACS point sources are detected outside the IRDC filaments. 

\begin{figure}
	\centering
	\includegraphics[width=\linewidth]{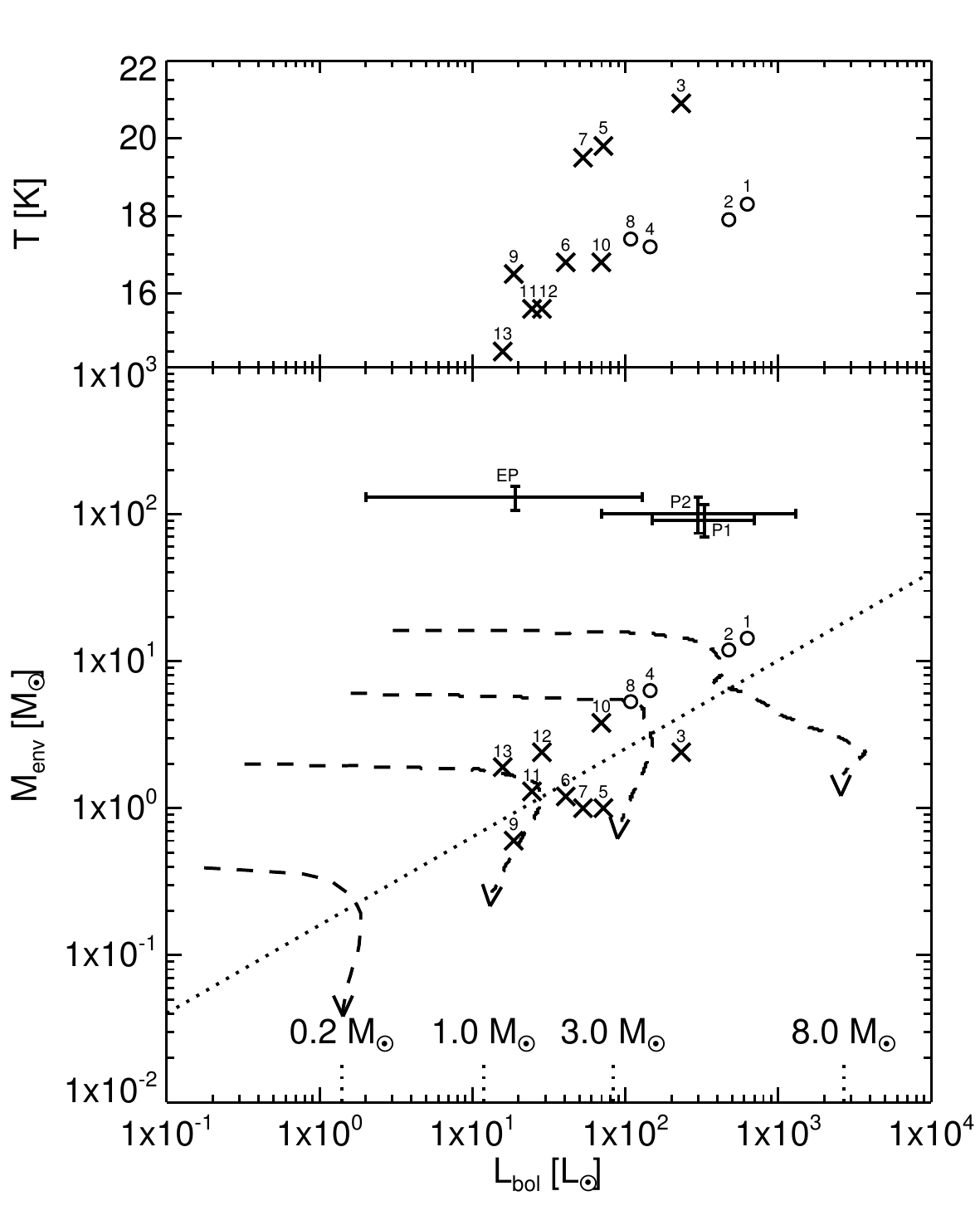}
 	\caption{The core temperatures (upper panel) and cold envelope mass (lower panel), derived from the SED fitting of the PACS point sources ($\lesssim 0.15\,$pc), are plotted against bolometric luminosities. Bolometric luminosities are derived from IRAC, MIPS and PACS bands. The numbers identify the point sources characterized in section \ref{sec:pointsrc}. The light-to-mass ratio derived from the sub-milliliter observation for different peaks (P1,P2 and EP) is plotted as well. In this observations the area of the traced mass reservoir is larger. 
 	The dashed lines represent the evolutionary tracks for YSO from \citet{Andre08}. The finale stellar masses (ZAMS) of those tracks are given above the lower axis.}
 	\label{fig:Lbol_vs_Menv}
\end{figure}

As in Section \ref{sec:Tmap}, we used a modified blackbody function to constrain the parameters of these point sources (Figure \ref{fig:blackbodies}), utilizing the dust model by \citet[thin ice mantles, $10^5$\,cm$^{-3}$]{Ossenkopf94}. The sources detected at 24\,$\mu$m cannot be fitted by a single component fit, since the warmer part of the envelope in the direct vicinity of the evolving protostar is contributing  significantly to the 24\,$\mu$m fluxes. Due to the rapidly increasing optical depth towards the innermost regions around the point-sources we only observe surface contributions of surrounding layers of warm dust. The calculations of mass and temperature assume optically thin material.

The 24\,$\mu$m data are highly sensitive to geometric effects in protostellar cores \citep[see][]{Dunham08,Ragan12}. A proto-star in which outflows have carved a cavity in its natal core will be detectable at 24\,$\mu$m if our line of sight is aligned with the cavity. 
In particular the shocked molecular hydrogen, visible as green structures in Figure \ref{fig:RGB_overview} (right, bottom panel), traces high velocity outflows. They originate from the rotational S9-S11 transition of excited \ce{H2} in the IRAC 4.5\,$\mu$m band. They are generally referred to as either ``green and fuzzy features'' \citep{Chambers09} or ``extended green  objects'' \citep[EGOs, ][]{Cyganowski08}. 
Because the 24\,$\mu$m flux arises primarily from the warm material immediately surrounding the proto-star (and not the cold outer region of the core), we do not include the 24\,$\mu$m flux in our SED fit to the PACS data but instead fit a second modified blackbody component (see Figure \ref{fig:blackbodies}). The best fit to the SED results from the sum to these two components. For Source 8, we used one modified blackbody function. The fit parameters are presented in Table \ref{tab:pnt_src}. 
 
The four most massive sources in our sample are located in the central part of the IRDC in the direct vicinity of the sub-millimeter peaks P1 and P2. They cover a mass range from 5.4 to 14.3\,M$_\odot$. From these sources, Source 8, is the only PACS source without a MIPS~24\,$\mu$m counterpart.
Source 1 and 2 are the brightest and most massive sources in the sample. They are associated with the SCUBA peak P1 and have a close separation of $\sim 12''$. Source 1 appears on top of a compact and diffuse background emission complex.  Only a  faint source  with 0.9\,M$_\odot$ is associated with the EP-SCUBA-peak of \citet{Ormel05}.
The point sources associated to P3 in the southern part of \G48 have only $\sim 1 -2$\,M$_\odot$. 

In Figure \ref{fig:Lbol_vs_Menv} we present the bolometric luminosities (IRAC to PACS),  the envelope masses ($\lesssim 0.15$\,pc), and the temperatures derived from cold components of the modified blackbody fits. 
In the luminosity-mass plot the dotted line follows the relation $M_{env}\sim L^{0.6}_{bol}$ and corresponds approximately to a threshold, where 50\% of the initial core mass %reservoir 
being converted into stellar mass $M_\star$ \citep[c.f.][]{Andre00,Bontemps96}. 
Sources with $M_{env}>M_\star$ and a high sub-millimeter to bolometric luminosities ratio $L_{\geq 350\mu m}/L_{bol} > 0.005$ are considered to be Class 0 \citep{Andre00}. 
Source 1 and 2 as well as Source 4 and 8 are detected as single SPIRE sources. But the fitted blackbodies indicate that the sub-millimeter fluxes should be equal for both sources in each case. For each source we calculate the sub-millimeter luminosity $L_{\lambda\geq 350\mu m}$ with 50\% of the corresponding SPIRE fluxes. All four sources mentioned before are good Class 0 candidates since $L_{\geq 350\mu m}/L_{bol}>0.005$ and $M_{env}>M_\star$. Source 1 and 2 have the highest envelope masses, and, as indicated by evolutionary tracks \citep[adopted from ][]{Andre08}, both could evolve into high-mass Zero Age Main Sequence stars (ZAMS) with $M_{ZAMS} \gtrsim 8\,\mathrm{M}_\odot$. 
These evolutionary tracks assume exponentially declining accretion rates. A recent study by \citet{Davies11}, comparing observations to various accretion models, showed that models with increasing accretion are in best agreement with the observational results. 
\citet{Molinari08} used a model with increasing accretion rates to create similar evolutionary tracks. According to these tracks Source 1 and 2 would only evolve into intermediate-mass stars ($<8\mathrm{M}_\odot$). However, it is still unclear on which spatial range a proto-star is accreting material. Source 1 and 2 are associated to the sub-millimeter peak P1. Due to the resolution of SCUBA compared to PACS a larger area is traced. Given the light-to-mass ratio for P1 derived by \citet{Ormel05}, still two massive stars ($M_{ZAMS} \gtrsim 8\,\mathrm{M}_\odot$) could form in the model from \citet{Molinari08}.
Assuming these sources are in a very young evolutionary stage, they are completely dominated by the accretion luminosity \citep{Krumholz07,Hosokawa09}. 

\newpage
\subsection{Molecular Line Emission} \label{sec:molecline}

To determine the parameters from the mapping line data, we used the GILDAS/CLASS \footnote{\url{http://www.iram.fr/IRAMFR/GILDAS/}} package.
As mentioned before we have line detections (CL$>99.73\%$) associated to the IRDC in the HCO$^+$, \ce{N2H^+}, HCN, HNC, $^{13}$CO and C$^{18}$O maps . While the line fitting for HCO$^+$, HNC and CO is done straightforward with a single Gaussian profile, N$_2$H$^+$ and HCN have a resolved hyperfine structure with multiple components.
For N$_2$H$^+$ a model profile containing seven hyperfine-structure components was fitted utilizing the GILDAS/HSF routine. The presented N$_2$H$^+$ maps containing the integrated line flux, line width and velocity correspond to the line center of the strongest component (93.17378\,GHz , ${J=1\rightarrow 0}$, ${F_1=2\rightarrow1}$, ${F=3\rightarrow 2}$).
The hyperfine-structure of HCN shows a triplet of components, and in contrast to N$_2$H$^+$, the relative line strength of each component varies for different IRDCs \citep{Afonso98,Vasyunina11}. 
Non-LTE effects especially in the 1--0 lines seem common \citep[see][]{Loughnane12}.
Given the noise level of the HCN map we could not retrieve an averaged profile. The fitting of the physical parameters is done on the strongest component.

\begin{deluxetable}{rcc}[th]
\tablecolumns{3}
\small
\tabletypesize{\footnotesize}
\tablewidth{0pt}
\tablecaption{Clumpfinder parameters}
\tablehead{
\colhead{Molecule} 	& \colhead{lower threshold$\,^\mathrm{(a)}$} 	& \colhead{step$\,^\mathrm{(a)}$} \\
\colhead{}			&  \colhead{\%}			& \colhead{\%}
}
\startdata 
N$_2$H$^+$ 		& 5.0		& 5.0	\\
HCO$^+$			& 5.8		& 5.8	\\
HCN		  		& 8.5		& 3.4	\\
HNC      		& 18.6		& 9.3	\\
$^{13}$CO (1-0)	& 35.2		& 2.3	\\
$^{13}$CO (2-1)	& 31.4		& 6.3	\\
C$^{18}$O (2-1)	& 8.8		& 5.5
\enddata
\label{tab:clumpfinder}
\tablecomments{$^\mathrm{(a)}$~The minimum detection thresholds and the contrast steps for \texttt{clumpfinder} are give in relation to the peak value in the corresponding map.}
\end{deluxetable}
 
The derived integrated line fluxes, velocities and line widths are compiled in Table \ref{tab:radio_lines}. This Table gives also the effective radius (see below), the total column density, the abundance and the traced mass.
\ce{HCO^+} and \ce{HCN} are good tracers for dense gas because of their high critical densities. \ce{N2H^+} has a lower critical density (see Table \ref{tab:line_prop}), but is not strongly effected by depletion from freeze-outs on grain surfaces. For CO the critical density is, depending on the transition, one or two orders of magnitudes lower  than for \ce{N2H^+}.
The total column density for a certain molecular species can be calculated with
\begin{equation}
N_u=\frac{8\pi A_{ul} k \nu^2}{h c^3}\int T_{mb}dv\,
\end{equation}
\begin{equation}
N_{tot}=\frac{N_uQ_{rot}(T)}{g_u}\exp(-E_u/T)
\end{equation}
where $A_{ul}$ is the Einstein coefficient, $\nu$ the rest frequency of the transition,
 $N_u$ and $E_u$ the upper state column density and energy,  $Q_{rot}$ the rotational partition function and $g_u$ the statistical weight of the upper level. The general physical parameters\footnote{The parameters can be obtained from the Leiden Atomic and Molecular Database (LAMBDA).\newline \url{http://www.strw.leidenuniv.nl/\~ moldata/}} for the observed lines are given in Table \ref{tab:line_prop}.
The abundance for a molecule is given in relation to the H$_2$ column density derived from the SED-fitted map from Section \ref{sec:Tmap}. 
The temperature estimates needed to calculate the different molecular abundances were obtained from the same SED-fitting.
The line emissions from \ce{HCO^+}, \ce{HCN}, \ce{HNC} and \ce{^{13}CO} can be optical thick, therefore the column densities and abundances provide only lower limits.

For the molecular line maps without any $3\sigma$ detection the upper limits for line intensity and column density, based on the noise estimates, are given in Table \ref{tab:radio_lines_upperlimits}.
 
\begin{deluxetable}{rcc}[th]
\tablecolumns{3}
\small
\tabletypesize{\footnotesize}
\tablewidth{0pt}
\tablecaption{Upper limits for non detected line species for MOPRA}
\tablehead{
\colhead{Line} 	& \colhead{line integral intensity$\;^{(a)}$} 	& \colhead{column density} \\
				& \colhead{[kms$^{-1}$K]}	& \colhead{[$10^{12}\,$cm$^{-1}$]} \\
}
\startdata 
CH$_3$C$_2$H 	& 0.50 &        \\
H$^{13}$CN		& 0.53 & 	3.8	\\
H$^{13}$CO$^+$  & 0.52 &   	0.7 \\
SiO             & 0.51 &   	1.2	\\    
C$_2$H         	& 0.52 &   		\\
HNCO            & 0.48 &   	6.4 \\
HCCCN           & 0.48 &   	2.0	\\
CH$_3$CN        & 0.50 &   	2.9 \\
$^{13}$CS       & 0.53 &   	5.9    
\enddata
\label{tab:radio_lines_upperlimits}
\tablenotetext{a}{
Upper limits are the integrated intensity a Gaussian-shaped line with a FWHM of 1\,kms$^{-1}$ would have at a peak intensity level of three times the RMS noise.}
\end{deluxetable}

Following \citet{Ragan06} the mass of each clump observed in the line maps is estimated using an assumed distance of 2.6\,kpc, an approximate size based on the extent of molecular emission, the molecular column densities, and approximate abundance calculated at the peak of absorption. 

In Figure \ref{fig:mopra} the SPIRE 250\,$\mu$m images are shown and overplotted by the different line intensity maps (\ce{HCO^+}, \ce{NH2^+}, \ce{HCN}, \ce{HNC} and \ce{CO}). We used the \texttt{clumpfind} \citep{Williams94} package to identify the different clumps in the molecular tracers. Four different clumps were found by clumpfinder, hereafter named C1-C4. The detection parameters for \texttt{clumpfinder} are given in Table \ref{tab:clumpfinder}.
C1, C2 and C3 are detected in HCO$^+$, HCN and \ce{N2H^+}. An additional clump C4 is detected in HCN.
In the $^{13}$CO (1-0) GRS map only C1 is detected. While C1 and C2 are detected in the HERA CO (2-1) maps and in HNC.
Because \texttt{clumpfind} does not make any assumption about the clump morphology an effective circular radius 
\begin{equation}
R_{eff}=\sqrt[]{A/\pi}
\label{eq:Reff}
\end{equation}
is given in Table \ref{tab:radio_lines}, where $A$ is the area fitted by \texttt{clumpfind}. 
Because the resolution of the HERA observations is three times better than MOPRA, the emission is less diluted and we therefore derive smaller $R_{eff}$.
All line maps show the strongest emission in the central part of the IRDC. The maps HCO$^+$, \ce{N2H^+}, HNC, and $^{13}$CO show an elongated structure in east-western orientation. HCN shows two distinct peaks (C1, C4); these peaks are shifted by $\sim 30''$ south-west of the compact SPIRE emission. 
All other MOPRA molecular tracers peak at C1, and while the maps appear shifted from the compact SPIRE emission, it is within the pointing error of MOPRA.
The peak positions for the HERA CO maps are much better aligned with the SPIRE peaks. 
In contrast to our observed HCO$^+$ (1-0) emission, the HCO$^+$ (3-2) observation from \citet{Ormel05} show multiple emission peaks. Their eastern emission peak is centered on the SCUBA clump, but the western emission is more diffuse.  With the given spatial resolution of MOPRA ($38''$) it is likely that we only observe a single integrated emission peak for HCO$^+$ from these multiple unresolved emitting sources. 

About $1\farcm3$ southwards of the central region another fainter emission peak (C2) can be found for HCO$^+$, HCN, HNC, \ce{N2H^+}, $^{13}$CO(2-1) and C$^{18}$O. It is not coincident with the compact SPIRE and sub-millimeter emission in this region. In the integrated line flux map for $^{13}$CO(1-0), no compact emission clump was found by \texttt{clumpfind}. The emission appears just as a diffuse extended lobe from the central peak. In the north-western part of the IRDC, where only diffuse emission appears in the SPIRE and sub-millimeter regime, the line data are much more noisy. For HCO$^+$, HCN and \ce{N2H^+} a compact emission peak (C3) can be identified, while $^{13}$CO shows a elongated diffuse emission lobe.

The three SCUBA clumps modeled by \citet{Ormel05} incorporate a mass reservoir of ${\sim 320\,\mathrm{M}_\odot}$. The masses derived for this region from the HCO$^+$, HCN, HNC, N$_2$H$^+$ and $^{13}$CO observations differ by up to 20\%. Given the uncertainties for the mass estimates, this is a good agreement.

\ce{CH3C2H} is a good temperature probe in dense molecular clouds \citep[e.g., ][]{Bergin94}. We detect four \ce{CH3C2H} transitions: $J=(9-8)$, $K=0,1,2,3$ with IRAM single pointing observations positioned on the sub-millimeter peak P1 (as indicated in the bottom left panel of Figure \ref{fig:NH2_Tmap}).
From the measured integral line intensities we derived the column densities for every particular transition using formulae 1-2 from  \citet{Bergin94}. Finally we computed the excitation diagram and estimated the excitation temperature and column density in the region. The resulting temperature is ${26\pm7\,\mathrm{K}}$ and the column density ${(3.0\pm 1.5)\cdot 10^{12}\,\mathrm{cm}^{-2}}$.
All necessary values, like $E_u$ - the upper state energy, $g_u$ - degeneracy and others, were taken from the CDMS data base \citep{Mueller01}

\begin{figure}
	\centering
	\includegraphics[width=\linewidth]{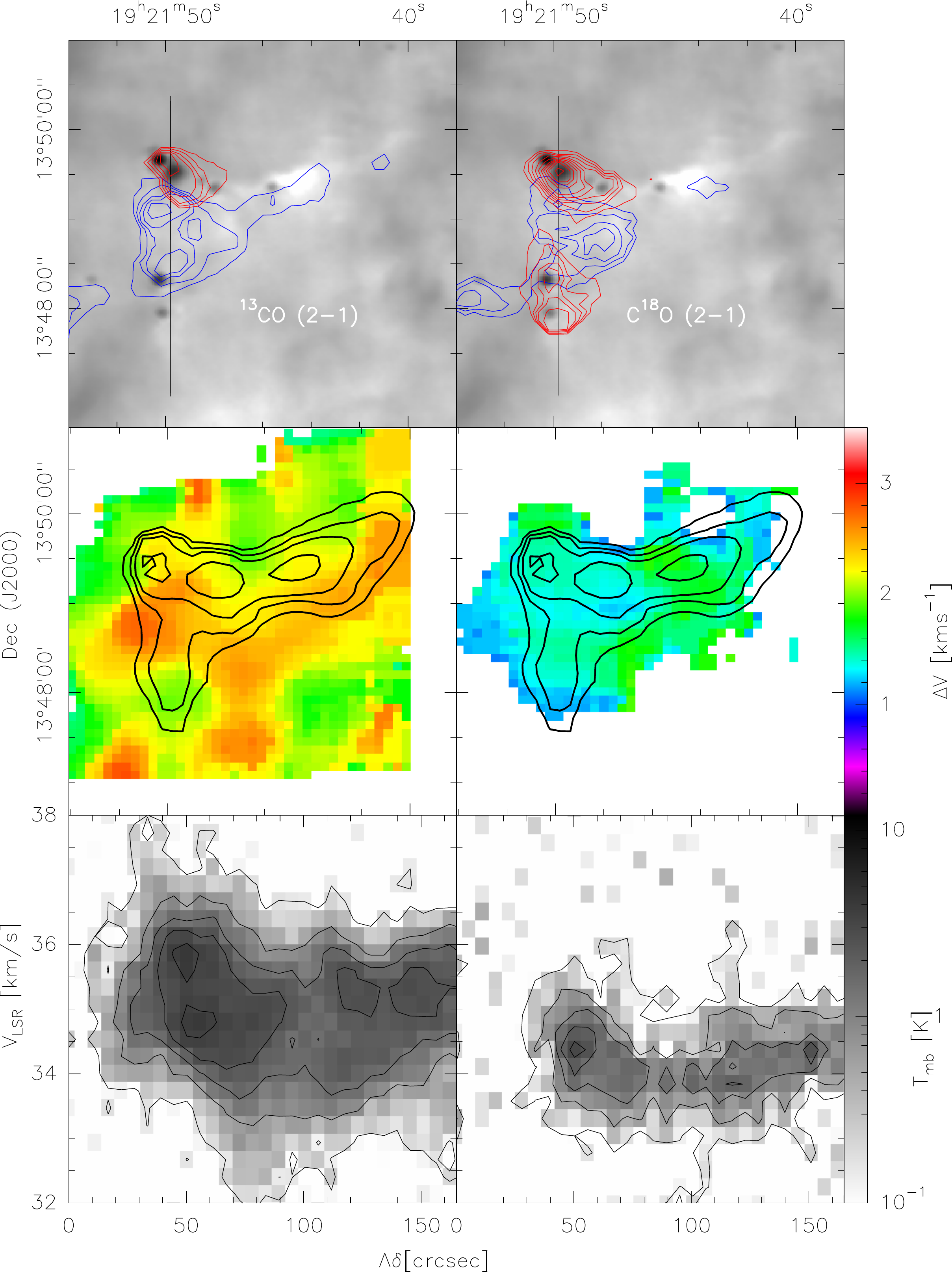}
	\caption{In the top panel the contours of the line intensities for $^{13}$CO (left) and C$^{18}$O (right) is plotted over the inversed PACS 70\,$\mu$m image. For $^{13}$CO (HERA) the blue contour is integrated between ${30.0-34.0\,\mathrm{kms}^{-1}}$ and the red contour between ${36.0-40.0\,\mathrm{kms}^{-1}}$. The integration in the C$^{18}$O HERA map is done over a more narrow range, due to higher noise level, with ${32.0-33.5\,\mathrm{kms}^{-1}}$ (blue) and ${34.5-36.0\,\mathrm{kms}^{-1}}$ (red). 
\newline	
	The middle plot shows the linewidth for the HERA CO data. The overplotted contours are the isotherms of 17.5, 18.5, 19.0, and 19.5\,K (see Figure \ref{fig:NH2_Tmap}).
\newline	
	The bottom plot shows the position-velocity diagram for $^{13}$CO and C$^{18}$O. The spatial orientation  of the p-v-plot is indicated in the top panels by a black vertical line.\vspace*{0.1cm}}
	\label{fig:HERA_CO}
\end{figure}
In Figure \ref{fig:HERA_CO} (bottom panel) the position-velocity (p-v) plots are shown for $^{13}$CO and C$^{18}$O (2-1). In the p-v plot for C$^{18}$O only a single velocity component is present. For $^{13}$CO two components are present at a 50$''$ offset, which refers to the spatial position of the Source 1 in the PACS points-source catalog and the sub-millimeter peak P1. Due to the small spread of just ${\sim1.5\,\mathrm{kms}^{-1}}$ it is questionable that this structure originates from an outflow cone. 
Other possible explanations could be a large rotational envelope or just two core-like dense CO structures along the line-of-sight. With the lack of additional observations with higher sensitivity and spatial resolution, no robust conclusion can be drawn.
The linewidth maps in Figure \ref{fig:HERA_CO} are fitted with a single Gaussian. A single Gaussian profile is also fitted there two components are found in the \ce{^{13}CO} map (Figure \ref{fig:HERA_CO}, middle plot). The resulting does not differ more then $0.5\,\mathrm{km\,s}^{-1}$ from a fit only on the strongest component.

With typical isotopic abundance ratios of ${[\ce{^{12}C}]/[\ce{^{13}C}]=60}$ and ${[\ce{^{16}O}]/[\ce{^{18}O}]=500}$ 
\citep{Wilson94,Lodders09} the HERA \ce{^{13}CO} and \ce{C^{18}O} abundances correspond to \ce{^{12}C^{16}O} abundances of  ${\sim1}-{1.5\cdot10^{-5}}$.
The expected CO abundance at a given Galactocentric distance ($D_\mathrm{GC}$) can be computed from the following relationship \citep[][and ref. within]{Fontani12}:
\begin{equation}
X^\mathrm{E}_\mathrm{C^{16}O}=8.5\cdot 10^{-5}\exp(1.105-0.13 D_\mathrm{GC} / D_\mathrm{kpc})
\label{eq:exp_CO_abun}
\end{equation}
For \G48 with $D_\mathrm{GC}=7.1$\,kpc this results in an expected CO abundance of ${X^E_\mathrm{C^{16}O}=1.0\cdot10^{-5}}$ thus consequently ${X^\mathrm{E}_\ce{C^{18}O}=2.0\cdot10^{-7}}$ and ${X^\mathrm{E}_\ce{^{13}CO}=1.7\cdot10^{-6}}$.
Comparing this expected values with the observed CO abundances gives the depletion factor:
\begin{equation}
f_D=\frac{X^\mathrm{E}_\ce{CO}}{X^\mathrm{obs}_\ce{CO}}
\label{eq:depletion_factor}
\end{equation}
Using the preceding isotopic abundance ratios results in average depletion factors of ${f_D^\ce{^{13}CO}=3.3}$ and ${f_D^\ce{C^{18}O}=3.7}$ for \G48.

We detected \ce{CO2} ice in the \textit{Spitzer} spectra. In Figure \ref{fig:spec} the \textit{Spitzer}/IRS spectrum for Source 2 is shown. 
The observed shape of the \ce{CO2} ice absorption feature at 15.2\,$\mu$m  is different for the source spectrum compared to the background spectra; the on-sources profile is significantly broader and different in shape compared to the nodded background profiles. This implies different ice temperatures and mixing ratios \citep[with e.g., \ce{H2O} and \ce{CH3OH}, see ][]{White09}.
The low spectral resolution  of the IRS/LL2 spectral order does not allow a proper fitting of temperature and mixing ratio to the laboratory data. Most likely the source \ce{CO2} profile originates from ice coated grains in the dust envelope of the embedded proto-star. The presence of \ce{CO2}-ice does not provide direct evidence for \ce{CO} freeze-out, since the conditions under which the surface conversion ${\ce{CO + OH ->[\text{surf}] CO2 + H}}$ takes place are still under debate \citep[e.g., ][]{Garrod11}.
In the gas-phase \ce{CO2} is predominantly produced by the following reaction
\begin{eqnarray*}
\ce{HCO + O -> CO_2 + H} \\
\ce{CO + OH -> CO_2 + H} \\
\ce{CO + O  -> CO_2 + H }
\end{eqnarray*}
To the last two reactions an activation energy barrier of 80\,K or higher is assigned \citep{Garrod11}. Therefore, gaseous \ce{CO2} creation becomes effective in cores which already have an internal heating source. 
The \ce{CO2} ice cannot be easily hydrogenated on dust grain surfaces though, so it will remain a terminal species of the CO evolution in lukewarm regions (Dmitry Semenov 2012, private communication).

\section{Discussion}\label{sec:conclusion}

\G48 is a quite isolated IRDC at a distance of 2.6\,kpc. Consequently, \G48 provides nonpareil observations of the collapse and fragmentation of a IRDC without the effects of external feedback, except for the diffuse ISRF.

We have derived line of
sight averaged \textit{Herschel} column density and temperature maps for this
source, and find minimum temperature of 16.8\,K, and outer  temperatures of
$\sim23\,$K. The corresponding column densities range from ${\sim 3\cdot 10^{22}\mathrm{cm}^{-2}}$ at in central part of \G48 to ${\sim 1\cdot 10^{22}\,\mathrm{cm}^{-2}}$ at the outer edges where our sensitivity drop sharply.
 
In this section we will discuss these results in combination with a wealth
of other observations with the goal placing this unique source in an
evolutionary context and determining its physical properties.

\subsection{Molecular line data, CO freeze out}
The comparison of the line data for \G48.66 with other IRDCs show results for \ce{N2H^+} which are similar to other IRDCs, while the \ce{HCO^+} results are different.
The average line width of \ce{HCO^+} is 2.5\,kms$^{-1}$ which is lower compared to other IRDCs, with typical values of ${\sim 3.2-3.3\,\mathrm{kms}^{-1}}$ \citep{Vasyunina11,Zhang11}. 
For \ce{N2H^+} we observed an average line width of 1.8\,kms$^{-1}$. Similar mean line widths were observed for other IRDCs with 1.8\,kms$^{-1}$ by \citet{Vasyunina11} and 2.0\,kms$^{-1}$ by \citet{Ragan06}. 
The mean column density of ${2.4\cdot 10^{12}\,\mathrm{cm}^{-2}}$ for \ce{N2H^+} in our sample is similar to the column density found for other IRDCs by \citet{Gibson09}. Comparing the amount of \ce{CO} freeze out one finds that the depletion factor is strongly dependent on the chemical age of the IRDC, hence the depletion varies drastically for different IRDCs. \citet{Hernandez11} and \citet{Hernandez12} reports a depletion factor of 5 for G035.39.-0033 which is a young IRDC of $\gtrsim7\cdot 10^4$\,yr similar to \G48. In contrast \citet{Miettinen11} found no depletion for their sample, and argued that this it is connected to a more evolved stage of their IRDCs where the increased star formation activity leads to sublimation of \ce{CO} from the grain surfaces.
\citet{Fontani12} reports higher values of $f_D=5-74$ for several southern IRDCs, the reason for this higher depletion factors is not completely clear, but might be explained by tracing denser regions of the cores by utilizing higher \C18O (3-2) transitions, compared to our \C18O (2-1) observations.
In the gaseous inter clump medium \ce{N2H^+} is destructed in the following reaction:
\ce{N2H^+ + CO -> HCO^+ + N2}  \citep{{Bergin07}}. 
The presence of \ce{N2H^+} is consistent with the low \ce{CO} abundance induced by the freeze-out on the dust-grain surfaces.

By comparing their \ce{HCO^+} observations to clump-models \citet{Ormel05} found an increase of turbulent motions towards the inner part of the clump. Their observations have higher resolution by a factor of two. Therefore, our observations are not good enough to trace down turbulence in the same way, but the increasing MOPRA linewidths inwards of the clumps are in agreement with the findings of \citet{Ormel05}. Our HERA \ce{CO} data have a higher resolution than their observations, but due to the low critical density of \ce{CO} we trace a larger column that might be dominated by envelope material also along the line of sight. Therefore, \ce{CO} does not qualify as a good turbulence tracer in dense environments as in our observations. 

%\newpage
\subsection{Cloud stability}
To assess the dynamical state of the molecular clumps we calculate the virial mass, using the line width $\Delta v$: 
\begin{equation}
M_\mathrm{virial}=k_2 R \Delta v^2
\end{equation}
whereas $k_2=126$ for a $n \sim 1/r^2$ profile \citep{MacLaren88},  $R$ is the effective radius of the molecular clumps, and $\Delta v$ is the FWHM of the respective line in $\rm km\,s^{-1}$. We also calculated the virial parameter $\alpha$ which is given by the ratio between the kinetically energy $E_\mathrm{kin}$ and the potential energy $E_\mathrm{pot}$:
\begin{equation}
\alpha=\frac{2E_\mathrm{kin}}{E_\mathrm{pot}}=\frac{M_\mathrm{vir}}{M}
\end{equation}
A value of $\alpha \approx 1$ represents a virial equilibrium, while $1< \alpha < 2$ is present for pressure confined clumps \citep{Bertoldi92}, $\alpha >2$ indicates a gravitationally unbound clump, and for $\alpha \ll 1$ the clump is collapsing. 
The virial mass and the virial parameter are presented in Table \ref{tab:radio_lines}. 
We see a large range, from less than 1 to over 7, in virial parameters depending on the species, making it difficult to assess the stability of each clump with certainty. 
However, while C1 seems to be off equilibrium and just confined by the surface pressure $(\alpha_\mathrm{C1}=1.9\pm0.7)$, C2 and C3 appear to be gravitationally unbound ($\alpha_\mathrm{C2}= 3.6\pm 1.3$ and $\alpha_\mathrm{C3}= 3.6\pm 0.4$, respectively). 

\begin{figure}
	\centering
	\includegraphics[width=\linewidth]{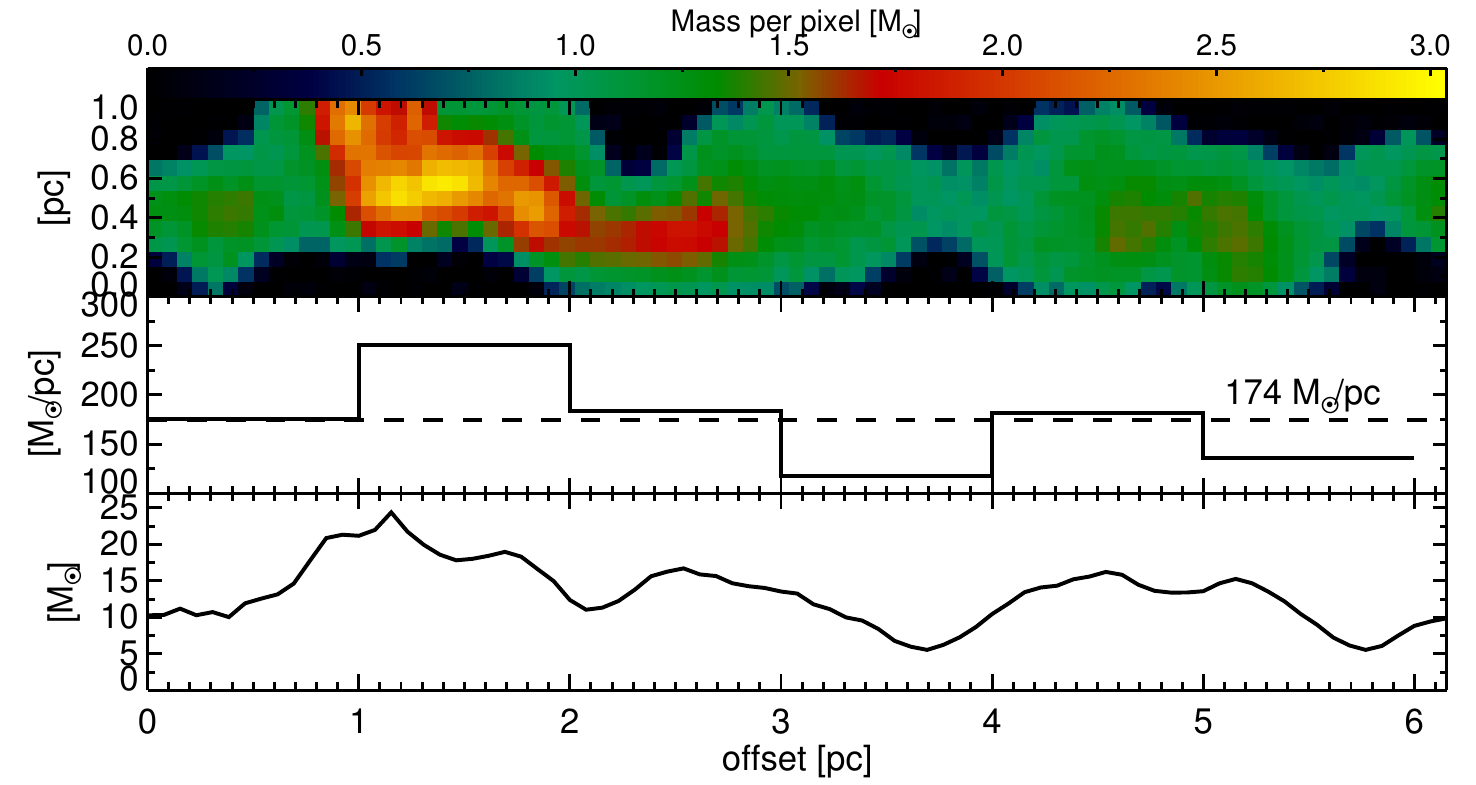}
		\caption{We present the line masses of the straightened filament of \G48. The upper panel shows the mass per pixel. The lower plot presents the masses summed over the width of the filament (hence the column of the upper plot). In the middle plot the masses are binned to 1\,pc. The average line mass for \G48 is given by the dashed line.}
	\label{fig:linemass}	
\end{figure}

Another crucial parameter is the virial line mass. An isothermal, nonmagnetic, self-gravitating filament in the absence of external pressure is in hydrodynamical equilibrium as long as the mass per unit length does not exceed $m_\mathrm{vir}=2\sigma^2/G$, where $\sigma$ is the velocity dispersion in the gas. If the filament is only thermally supported the velocity dispersion is given by the thermal sound speed $c_s$ \citep{Ostriker64}, while in the case of additional turbulence support the velocity dispersion is given by $\sigma=\Delta v / \sqrt{8 \ln 2}$ \citep{Fiege00}, where $\Delta v$ is the velocity dispersion. The critical line mass for a thermally supercritical filament in \G48 is $M^\mathrm{therm}_\mathrm{line}=24$\,M$_\odot/\mathrm{pc}$, whereas the critical line mass calculated by the mean linewidth of \G48, to account also for turbulent motions, is $M^\mathrm{crit}_\mathrm{line}=71$\,M$_\odot/\mathrm{pc}$. 
To highlight the derived linemass, we straightened the filament of \G48 utilizing a set of non-uniform cubic splines \citep{Kocsis91} and calculated the mass in each pixel. Binning the pixels results in an average line mass of 174\,M$_\odot$/pc (see Figure \ref{fig:linemass}). 
The observed line masses are $\sim 2.5$ larger than the critical line mass, indicating that the \G48 filament does not contain enough internal support to counter the collapse while the contributions from magnetic fields are expected to be negligible (see below). Line masses observed for other IRDC seem to exceed $M^\mathrm{therm}_\mathrm{line}$. 
However, for strongly fragmented cores with significantly lower mass in Taurus line masses similar to $M^\mathrm{therm}_\mathrm{line}$, hence predominately thermal support, were observed \citep{Schmalzl10}.
In contrast to \G48 several IRDCs show line masses below or similar to $M^\mathrm{crit}_\mathrm{line}$. This cannot be explained by the evolutionary stage of the IRDCs alone. The young and starless Coalsack cloud  and the adjacent G304.74+01.32 show line masses of  less or equal to $M^\mathrm{crit}_\mathrm{line}$ \citep[][ respectively]{Beuther11,Miettinen12}. 
Also \citet{Hernandez12} found the young H filament of G035.39--0.33 to be virial undercritical, although these results rely several uncertain morphological assumptions about the IRDC. 
\G48 and the Nessie IRDC are more evolved, showing signs of more active star formation such as ``green and fuzzy'' features, but the latter region still shows line widths a factor of 5 below the critical value for turbulence support \citep{Jackson10}. Furthermore the presence of magnetic fields does not change the critical line mass significantly \citep{Fiege00}. However, the higher line mass of 385\,M$_\odot$ for the more evolved Orion~A cloud \citep{Jackson10} supports the view of increasing line masses along the evolutionary sequence.

\subsection{Fragmentation and point source separation}
In the central part of the IRDC a concentration of point sources is found. The projected median separation of PACS-point-sources is $\sim 0.3$\,pc, which is approximately the average separation in the central part of \G48.
This separation is about three times smaller than in other IRDCs\footnote{The separation is about 0.9\,pc for the IRDCs  G011.11--0.12 \citep{Henning10} and G049.40--0.01 \citep{Kang11}; and 0.75\,pc for G304.74--1.32 \citep{Miettinen12}.}.
The median separation in the  R1 fragment of the Coalsack dark cloud is even an order of magnitude higher compared to \G48 \citep{Beuther11}, although here this study uses gas/dust peak separations instead of point source separation. The key question is how to explain these different values. One could assume that we just suffer from a projection effect in \G48. Considering we see an elongated filament and given the observed morphology it is a feasible assumption that \G48 is aligned approximately face-on towards the line-of-sight. Even if one would assume that \G48 is aligned off by $\pi/4$ this would just account for a factor of $\sqrt{2}$. 
In contrast to \G48 the mass reservoir in the IRDCs mentioned above is significantly higher; these other dark clouds do not share the isolated nature of \G48. 
While external feedback and the overall evolutionary stage of the parent cloud might have an important impact on the fragmentation, evaluation of the entire high-mass part of the EPoS sample in \citet{Ragan12} show that the average point source separation is not correlated with the cloud mass (Pitann et. al, in prep.). 
Recent molecular line observation of G28.34+0.06 found a separation of just 0.16\,pc \citep{Wang11} to 0.19\,pc \citep{Zhang09}. The interferometric data analyzed in \citet{Wang11} show substructures on a $\sim0.01$\,pc scale inside the cores. 

Given the wide spread of source separation it should be discussed whether \G48 exhibits the same mode of collapse as the IRDCs with larger separation or if some substantially different physical processes can account for the smaller separation.
To address this question we compare the observational results to analytical considerations.
A first simple analysis is to compare the source separation to the Jeans length.
The separation in \G48 is still several times larger than the thermal Jeans length ($\lambda_\mathrm{J}\approx 0.08\,\mathrm{pc}$ at $10^5\,\mathrm{cm}^{-3}$, 19\,K). Therefore, a thermal Jeans instability is not the dominant fragmentation mode. Fragmentation below this spatial scale has been observed for several massive cores \citep[e.g., ][]{Hennemann09,Wang11}.

A way to assess the spatial separation scale is to consider the collapse of a cylinder (with a radius $R_c$) of an incompressible fluid. It is predicted that several cores form, periodically placed at a characteristic length scale $\lambda_\mathrm{frag}=R_c$ \citep{Chandrasekhar53}. 
For an infinite cylinder of isothermal self-gravitating gas the characteristic fragmentation scale is $\lambda_\mathrm{frag}=22H$, where $H$ is the isothermal scale height $H=c_s/(4\pi/G\rho_c)^{1/2}$ and $c_s$ the thermal sound speed, $G$ the gravitational constant and $\rho_c$ the central density in the cylinder \citep{Nagasawa87}. When a isothermal cylinder with a given radius is embedded into a uniform medium the separation is  $\lambda_\mathrm{frag}\approx 22H$ for $R\gg H$, but for the opposite case $R\ll H$ the length scale is  $\lambda_\mathrm{frag}\approx 11 R$ \citep{Jackson10}. Given the measured temperatures the typical thermal sound speed is $c_s=0.23$\,kms$^{-1}$ in \G48. With $0.5-0.8$\,pc the width of the filament is much larger than the scale height. The calculated length scale $\lambda_\mathrm{frag}=22H\approx0.3$\,pc (with $n=10^5$\,cm$^{-3}$, $c_s=0.23$\,kms$^{-1}$) is the same as the measured separation in \G48. However, $\lambda_\mathrm{frag}$ is affected by uncertainties, since the real value of the central density is unknown. Moreover, the most regular positioning of sources occur in the central part of the IRDC along a clearly changing temperature profile which is a distinct derivation from the isothermal assumption.

\subsection{Filament width}
The typical filament width seen in Figure \ref{fig:linemass} is $0.5-0.8$\,pc. \citet{Peretto12} and \citet{Arzoumanian11} find typical filament widths of $0.08-0.1$\,pc in two different  low-mass star-forming regions in the Gould-Belt \textit{Herschel} GTO project. These widths are a factor $\sim 3$ smaller compared to the SPIRE 350\,$\mu$m. IRAC provides a much better resolution, but using the IRAC column density contours shown in Figure \ref{fig:NH2_Tmap}, results in an estimated filament width of $0.15-0.27$\,pc. Because this widths a based on the column density threshold, they might be larger than the corresponding FWHM of a fitted Gaussian profile.  \citet{Arzoumanian11} calculated the line mass for their filaments. All filaments have line mass below the average line mass for \G48 and most of the filaments are in contrast to \G48 below the critical thermal line mass. Also most of the filaments in the Pipe Nebula are thermally subcritical \citep{Peretto12}.
These results might indicate a trend of larger filament widths towards supercritcal filaments (with respect to turbulence).

\subsection{Temperature profiles} \label{sec:temp_disc}
Recent studies have shown that the temperature structure of star-forming regions is non-isothermal. Dust temperature maps for low-mass star-forming Bok-globules are shown in, e.g., \citet[CB244]{Stutz10}.
Preliminary results for IRDCs were shown by \citet{Peretto10} for G030.49+0.96 using SED-fitting. 
The minima in their temperature maps are aligned with the column density peaks as in \G48 and the subsequent studies. The minimum temperatures for G030.49+0.96 are lower compared to \G48 and the two following studies.
The work of \citet{Battersby11} and \citet{Beuther12} show similar temperature ranges as for \G48.  While \citet{Battersby11} provided a large scale temperature and column density map, \citet{Beuther12} present a more detailed study of IRDC~18454 adjacent to the mini-starburst W43. In contrast to IRDC~18454, \G48 provides a temperature map of an externally undisturbed filament with the better resolution of the convolution to the SPIRE~350\,$\mu$m beam (as for G030.49+0.96, IRDC~18454 data were convolved to the SPIRE 500\,$\mu$m beam). The low temperatures in the cloud centers indicate that they are very young structures. 

\citet{Wilcock12} used a different approach. They compared radiative transfer models of starless purely
externally heated cores with the observed embedded cores in IRDCs in the \textit{Spitzer} and \textit{Herschel} bands. They found core temperature profiles with minimum temperatures of $8-11\,$K which increase by $\sim 10-17\,$K towards the edge of the core. 
While our temperature mapping only observe an average line-of-sight temperature of the optically thin material, the RT model can retrieve temperature profiles along the line of sight. Since these profiles probing also the optically thick regime, the minimal temperatures are obviously lower.
On the other hand, the RT models require an assumption about the geometry of the cores and cannot reassemble the diffuse temperature structure of the whole IRDC.

The temperature profile for \G48 shows a drop towards the inner part of the detected sub-millimeter peaks. Theoretical considerations show that gas-phase coolants, as C$^+$  and CO, are efficient for low densities, while the energy transfer between gas and dust is the dominant cooling process for higher densities \citep[e.g.,][]{Goldsmith01,Glover12}. 
With the presence of IRAC and MIPS 24\,$\mu$m point sources one would expect to see the imprint of this embedded heating sources in the temperature maps. The SPIRE 350\,$\mu$m beam size limits the resolution of our temperature maps to $\sim 0.3\,pc$ so that the embedded heating sources remain unresolved.
The dust temperature mapping as presented in this work can provide a versatile tool to test the predicted coupling between gas and dust. 
Methylacetylene (\ce{CH3C#CH}) can be used to determine gas temperatures in massive star-forming regions \citep[cf.][]{Miettinen06}. 
IRAM pointed observations of \ce{CH3CCH} towards the peak position of the sub-millimeter peak P1 results in a gas temperature of $26\pm7$\,K. Although this gas temperature is 7\,K higher than the dust temperature the difference is not significant due to the high uncertainty in the gas temperature estimate. 
However the trend to higher gas temperatures is a good motivation for further observations of \ce{CH3C#CH} or \ce{NH3}. Especially gas temperature mapping can provide useful information on the gas-dust coupling at higher densities in molecular clumps.
Future work should study the relation between dust temperature structures and gas temperature maps as provided by ammonia \citep[as in][]{Pillai06,Ragan11} or methylacetylene line observations. Such data should provide further observational evidence for the predicted strong thermal coupling between gas and dust at higher densities.

The northern and southern extent of the IRDC (hosting C2 and C3) show a much flatter temperature profile with higher temperatures compared to the central part of the filament. Furthermore, the column densities are lower towards C2 and C3. This leads to the conclusion that the gas and dust densities are lower in these parts of the \G48. Pure isothermal cylindrical collapse can hardly explain this density gradient.
\citet{Hartmann07} showed by modeling the density structures for the Orion A cloud that, even in the isothermal case, a simple initial arrangement of a slightly rotating gas sheet can result in filamentary structures with complex density structures. 

As in the studies mentioned above, it is a clear finding of this work that the dust temperatures are non-isothermally distributed in an IRDC. What impact could the modeling of thermal profiles have?
Thermal line broadening is significantly smaller compared to the observed linewidths. Therefore, a change of local temperatures should not have a notable influence on the turbulence structure inside the core. On the other hand the thermal profile towards the core can play an important role in the support against fragmentation, since the Jeans mass 
is strongly dependent on temperature as $M_\mathrm{J}\sim T^{3/2}\rho^{-1/2}$. 
Theoretical studies of clustered star formation utilizing polytropic equations of states $P\sim \rho^\gamma$ and temperature dependent density profiles $T \sim \rho^{\gamma-1}$ indicate that the number of fragments in a cloud is strongly dependent on the polytropic index $\gamma$ \citep{Li03}. For the non-isothermal case ($\gamma\neq 1$) the polytropic index reflects the balance between cooling and heating. 
Deriving dust temperatures maps as in this work is the first step towards a better understanding of spatial distributed cooling and heating inside the core. A more detailed discussion of thermal properties can be found in e.g., \citet{Larson05} and \citet{Commercon11}.

\section{Summary} \label{sec:summary}
\noindent The following gives a brief summary of the results presented in this paper:
\begin{itemize}
\item \G48 is a quite isolated, mono-filamentary IRDC. The GMC W51 observed in the same field is twice as distant as \G48.
\item The dark filament observed by \textit{Spitzer} does not appear as a large coherent extinction silhouette in the PACS bands. Only the region of highest extinction in the central part is seen in extinction. The filament is clearly seen in emission in the SPIRE bands beyond 250\,$\mu$m. This is in good agreement with the sub-millimeter emission maps from SCUBA and ATLASGAL.
\item We present reliable temperature and column density maps based on pixel-by-pixel SED fitting of \textit{Herschel}, ATLASGAL and SCUBA legacy maps. Regions of low emissions outside of the galactic plane were utilized to estimate the global background level for the \textit{Herschel} maps.
The location of the temperature  minima are well aligned with the sub-millimeter and column density peaks.  The temperatures drop to 17.5\,K in the central part of \G48. The heating by embedded MIPS 24\,$\mu$m point sources is not detected, due to beam dilution.
\item We derived a total mass of 390\,M\sun for \G48 by applying a threshold of ${1.1\cdot 10^{22}\,\mathrm{cm}^{-2}}$ to the  SED fitted column density maps.
\item ``Green and fuzzy'' features reveal the central parts of the IRDC as the sites of most active ongoing star formation.
\item We detected 13 point sources in all three PACS bands. Only one source lacks a MIPS 24\,$\mu$m counterpart. 
\item SED fitting to these sources result in core masses of $0.6-14.3$\,M\sun, while the bolometric luminosity ranges between 15 and 630\,L\sun. The two most massive cores (Source 1 and 2) have mass of 14.3 and 11.9\,M$_\odot$ and a bolometric luminosity of 630 and 475\,L\sun, respectively. Comparing their light-to-mass ratio to evolutionary models, both sources are likely to evolve into high-mass stars with $M\gtrsim8\,\mathrm{M}_\odot$. Applying the definition from low-mass star formation, four cores are good candidates for ``Class~0'' objects.
\item The average line mass of the \G48 filament is 174\,M\sun pc$^{-1}$. Compared to theoretical predictions the whole filament is collapsing, even under the assumption of internal turbulent support.
\item The point source separation in \G48 is significantly smaller than in other IRDCs, but in good agreement with the predictions by the isothermal cylindrical collapse model. 
\item According to our virial analysis the molecular clumps found in \G48 are gravitationally unbound or just pressure confined. 
\item Our observations are consistent with previous observations by \citet{Ormel05}, which deduced increasing turbulent motion towards the inner parts of the clumps.
\item We detected several molecular species, such as \ce{HCO^+}, \ce{HCN}, \ce{HNC}, \ce{N2H^+}, \ce{^{13}CO}, \ce{C^{18}O} (and \ce{CH3CCH}). The mapping observation showed several molecular clumps associated to the sub-millimeter peaks.
\item We found CO depletion towards the molecular clumps C1 + C2 with factor of $f_D \approx 3.5$. 
\item The low abundance of CO towards the sub-millimeter peaks P1 and P2 permit relative high abundances of \ce{N2H^+}. 
\end{itemize}

\acknowledgements
\begin{scriptsize}

\noindent \textit{Acknowledgments}

We thank Zoltan Balog and Marc-Andre Bessel for their help with the optimal reduction of the IRAC/GLIMPSE and MIPSGAL maps. We appreciate the help of Bernhard Sturm with the polynomial background fitting of the \textit{Spitzer}/IRS spectra. We are grateful to Dmitry Semenov for the enlightening discussions regarding astro-chemical problems. \\
This work is based in part on observations made with the \textit{Spitzer Space Telescope}, which is operated by the Jet Propulsion Laboratory, California Institute of Technology, under a contract with NASA. 
This research made use of Tiny Tim/Spitzer, developed by John Krist for the Spitzer Science Center. The Center is managed by the California Institute of Technology under a contract with NASA. 
This paper is using data from the Mopra radio telescope. The Mopra radio telescope is part of the Australia Telescope National Facility which is funded by the Commonwealth of Australia for operation as a National Facility managed by CSIRO. 
This publication makes use of molecular line data from the Boston University-FCRAO Galactic Ring Survey (GRS). The GRS is a joint project of Boston University and Five College Radio Astronomy Observatory, funded by the National Science Foundation under grants AST-9800334, AST-0098562, \& AST-0100793. This research has made use of NASA's Astrophysics Data System. 
This research has made use of the SIMBAD database, operated at CDS, Strasbourg, France. 
This research made use of APLpy, an open-source plotting package for Python hosted at \url{http://aplpy.github.com}
The work of S.R. and A.M.S.  was supported by the Deutsche Forschungsgemeinschaft
priority program 1573 ("Physics of the Interstellar Medium").
T.V. wishes to thank the National Science Foundation (US) for its funding of the astrochemistry program at the University of Virginia.
\end{scriptsize}

{}

\clearpage
%\begin{landscape}
\begin{turnpage}
\begin{deluxetable}{cllcc ccccc cc}
\tablecolumns{13}
%\small
%\rotate
%\tabletypesize{\tiny}
%\tabletypesize{\footnotesize}
\tabletypesize{\small}
\tablewidth{0pt}
\tablecaption{Molecular line observations}
\tablehead{
\colhead{Src}				& \colhead{R.A. (2000)}			& \colhead{Dec (2000)}   		& \colhead{$R_\mathrm{eff}\;^\mathrm{(a)}$}		& \colhead{$V_{LSR}\;^\mathrm{(b)}$}				&\colhead{$\Delta v\;^\mathrm{(c)}$} 		 & \colhead{$\int T_{mb} dv\;^\mathrm{(d)}$}  			& \colhead{$N(X)\;^\mathrm{(e)}$}	&  \colhead{Abundace$\;^\mathrm{(f)}$}  & \colhead{$M(X)\;^\mathrm{(g)}$}  & \colhead{$M_\mathrm{vir}\;^\mathrm{(h)}$} & \colhead{$\alpha_\mathrm{vir}\;^\mathrm{(i)}$} \\
								& \colhead{$[$hh:mm:ss$]$}		& \colhead{$[$dd:mm:ss$]$}   	& \colhead{arcsec}  & \colhead{km$\cdot$s$^{-1}$}	& \colhead{km$\cdot$s$^{-1}$} & \colhead{K$\cdot$km$\cdot$s$^{-1}$} & \colhead{$10^{12}$cm$^{-2}$}	& \colhead{$\frac{N(X)}{N(\mathtt{H_2})}\cdot 10^{-10}$}	& \colhead{$[\mathrm{M}\sun]$} & \colhead{$[\mathrm{M}\sun]$} & \colhead{}  }
\startdata 
\multicolumn{2}{l}{HCO$^+$} & & & & & & & & & \\
C1		& 19:21:48.958		& $+$13:49:17.04	& 60	& 33.90~(0.16)	& 2.50~(0.11)	& 3.84	& $>5.0$	& $>2.0$ 	& 310	& 605 	& 1.95	\\
C2		& 19:21:51.016		& $+$13:47:47.03	& 50	& 33.92~(0.12)	& 2.06~(0.15)	& 1.64	& $>2.3$	& $>1.6$ 	& 100	& 343	& 3.43	\\
C3		& 19:21:35.569		& $+$13:52:02.04	& 45	& 34.20~(0.16)	& 2.99~(0.26)	& 1.40	& $>1.6$	& $>1.2$	& 150	& 649	& 4.33	\\
\hline
\multicolumn{2}{l}{HCN} & & & & & & & & & \\
C1		& 19:21:48.958		& $+$13:49:17.04	& 40	& 34.0~(0.4)	& 3.6~(0.7)		& 1.29	& $>8.9$	& $>3.5$	& 230	& 837	& 3.64	\\
C2		& 19:21:51.016		& $+$13:47:47.03	& 42	& 33.8~(0.3)	& 2.3~(0.5)		& 0.84	& $>6.1$	& $>4.3$ 	& 50	& 359	& 7.18  \\
C3		& 19:21:35.569		& $+$13:52:02.04	& 40	& 33.7~(0.4)	& 2.5~(0.5)		& 1.04	& $>7.5$	& $>4.9$	& 115	& 403	& 3.50  \\
C4		& 19:21:46.898		& $+$13:49:17.04	& 41	& 34.4~(0.5)	& 2.6~(0.4)		& 1.04	& $>7.2$	& $>3.7$ 	& 165	& 448	& 2.72	\\
\hline
\multicolumn{2}{l}{HNC} & & & & & & & & & \\
C1		& 19:21:48.958		& $+$13:49:17.04	& 63	& 33.81~(0.11)	& 2.50~(0.21)	& 1.27	& $>2.7$	& $>1.1$	& 450	& 636 	& 1.41	\\
C2		& 19:21:51.016		& $+$13:47:47.03	& 47	& 33.77~(0.12)	& 1.66~(0.18)	& 2.01	& $>4.5$	& $>3.1$	& 95	& 209	& 2.20	\\
C3		& 19:21:35.569		& $+$13:52:02.04	& 39	& 33.91~(0.28)	& 2.4~(0.4)		& 0.83	& $>1.8$	& $>1.2$	& 140	& 363	& 2.59	\\
\hline
\multicolumn{2}{l}{N$_2$H$^+$} & & & & & & & & & \\
C1		& 19:21:48.958		& $+$13:49:17.04	& 72	& 33.60~(0.17)	& 1.47~(0.25)	& 4.69	& 68.7		& 27.0 		& 370	& 251	& 0.68	\\
C2		& 19:21:51.016		& $+$13:47:47.03	& 71	& 33.59~(0.11)	& 1.10~(0.22)	& 2.92	& 45.0		& 31.5 		& 85	& 139	& 1.64	\\
C3		& 19:21:35.569		& $+$13:52:02.04	& 50	& 33.8~(0.5)	& 2.9~(0.7)		& 2.06	& 31.4		& 20.5  	& 165	& 679	& 4.15	\\
\hline
\multicolumn{2}{l}{$^{13}$CO (1-0)} & & & & & & & & & \\
C1		& 19:21:48.449	& $+$13:49:17.90	& 76	& 33.72~(0.11)	& 2.11~(0.03)	& 8.53	& $>2.0\cdot10^4$ 	& $>7.9\cdot 10^3$ 	& 380	& 546	& 1.44 \\
\hline
\multicolumn{2}{l}{$^{13}$CO (2-1)} & & & & & & & & & \\
C1			& 19:21:48.449		& $+$13:49:17.90	& 19  	& 35.1~(0.22)	& 2.53~(0.06)		&	14.9	& $>1.6\cdot10^4$ 	&  $>6.3\cdot 10^3$	& 60	& 196	& 3.27	\\
C2			& 19:21:51.016		& $+$13:47:47.03	& 16   	& 34.9~(0.3)	& 2.49~(0.06)		&	 5.9	& $>6.4\cdot10^3$	&  $>4.5\cdot 10^3$ & 40	& 160	& 4.00 \\
 & & & & & & & & & & \\
%\hline
\multicolumn{2}{l}{C$^{18}$O (2-1)} & & & & & & & & & \\
C1			& 19:21:48.449		& $+$13:49:17.90	& 18   	& 34.2~(0.3)	& 1.54~(0.11)	&	2.3		& $1.2\cdot10^3$ 	&  $4.9\cdot10^2$		& 55	& 69	& 1.25  \\
C2			& 19:21:51.016		& $+$13:47:47.03	& 15   	& 34.02~(0.20)	& 1.88~(0.15)	&	1.6		& $8.6\cdot10^2$ 	&  $6.0\cdot 10^2$		& 30	& 86	& 2.87
 \enddata
\label{tab:radio_lines}
\tablecomments{$^{(a)}$~Effective radius from \texttt{clumpfinder}, $^{(b)}$~rest frame velocities, $^{(c)}$~line width, $^{(d)}$~integrated line intensities, $^{(e)}$~column density of the observed molecular species, $^{(f)}$~abundance of the molecular species with respect to H$_2$, $N_\mathrm{H_2}$ is calculated from the 870\,$\mu$m data (and 8\,$\mu$m extinctions in brackets), $^{(g)}$~mass estimates from molecular line maps, see text, $^{(h)}$~virial mass $M_\mathrm{vir}=126R_\mathrm{eff}\Delta v^2$, see text, $^{(i)}$~virial parameter $\alpha=M_\mathrm{vir}/M$.}
\end{deluxetable}
\end{turnpage}
%\clearpage
%\end{landscape}
\end{document}